\documentclass[reviewcopy]{elsarticle}

\usepackage[reviewcopy]{adndt}
\usepackage{longtable}

\usepackage{mathptmx}

\usepackage{amsmath}

\usepackage{amssymb}

\biboptions{square,sort&compress}
\bibpunct[]{[}{]}{,}{n}{}{;}
\begin{document}

\begin{frontmatter}

\journal{Atomic Data and Nuclear Data Tables}


\title{Comment on ``Atomic structure and electron impact excitation of Al-like ions (Ga--Br)" by HB Wang and G Jiang  in
At. Data Nucl. Data Tables 148 (2022) 101532}

  \author[One]{Kanti M. Aggarwal\corref{cor1}}
  
  \ead{K.Aggarwal@qub.ac.uk}
 
   \author[One]{Ken W. Smith}

  \cortext[cor1]{Corresponding author.}
  \fntext[Y]{{\em E-mail address}: K.W.Smith@qub.ac.uk}

  \address[One]{Astrophysics Research Centre, School of Mathematics and Physics, Queen's University Belfast,\\Belfast BT7 1NN,
Northern Ireland, UK}

\date{06/05/2016} 

\begin{abstract}  
In a recent paper,  Wang and Jiang (At. Data Nucl. Data Tables 148 (2022) 101532) have reported data for energy levels, radiative rates (A-values), and effective collision strengths ($\Upsilon$) for some transitions of five Al-like  ions, namely Ga~XIX, Ge~XX, As~XXI, Se~XXII, and Br~XXIII. On a closer examination we find that their reported data for energy levels and A-values are generally correct, but not for  $\Upsilon$. Their  $\Upsilon$ values, for all transitions (allowed or forbidden) and for all ions,  invariably decrease at higher temperatures. This is mainly because they  have  adopted a limited range of  electron energies for the calculations of collision strengths. We demonstrate this with our calculations  with the Flexible Atomic Code (FAC), and conclude that their $\Upsilon$ values are inaccurate, unreliable, and should not be adopted in any applications or modelling analysis. \\

Received 28 April 2023; Accepted 17 May 2023 \\


\end{abstract}

\end{frontmatter}




\newpage

\tableofcontents
\listofDtables
\listofDfigures
\vskip5pc


\section{Introduction}
\label{intro}

Atomic data, including energy levels, radiative rates (A-values), collision strengths ($\Omega$), and effective collision strengths ($\Upsilon$) are required for a wide range of ions for applications and modelling of a variety of plasmas, such as solar, astrophysical and fusion. For many ions such data are reliably available in the literature, but for some are not. To fulfil the gap, Wang and Jiang \cite{wj} have recently reported data for these parameters for five Al-like ions, namely Ga~XX, Ge~XX, As~XXI, Se~XXII, and Br~XXIII. For the calculations of energy levels and A-values, they have adopted the GRASP (general-purpose relativistic atomic structure package) code. This is a fully relativistic code and was  originally  developed by  Grant et al. \cite{grasp0}, but  has since undergone through multiple revisions. For the most recent version and its manual, see J{\"o}nsson et al. \cite{grasp}.

For the calculations of collisional data, i.e. $\Omega$ and $\Upsilon$, Wang and Jiang \cite{wj}  have adopted the Dirac atomic  $R$-matrix code (DARC). This is also a fully relativistic code, is compatible with GRASP, and was developed by P.H. Norrington and I.P. Grant. The code has never  been published but is freely  available at the website: {\tt http://amdpp.phys.strath.ac.uk/UK\_APAP/codes.html}. Furthermore, they have resolved resonances for the calculations of $\Omega$, and subsequently of $\Upsilon$. The resonances often significantly contribute to the determination of $\Upsilon$ values, particularly for the forbidden transitions, and towards the lower end of the temperature ranges. Therefore, in principle, their calculations should be the most accurate available to date. Unfortunately, this is not true as discussed below.

 Wang and Jiang \cite{wj} have not reported numerical data for $\Omega$, except showing graphically for a few transitions in their figs. 3--5, and that too in a very limited energy range below 140~Ryd. However, in their tables~11--15 they have reported $\Upsilon$ values for transitions from the lowest two levels ($^2$P$^o_{1/2}$ and $^2$P$^o_{3/2}$) of the ground configuration 3s$^2$3p  to higher excited levels, up to 30 -- see Table~A for the specific levels.  The $\Upsilon$ values are listed over a wide range of temperatures, up to 10$^8$~K, but {\em decrease} for all transitions towards the higher end of T$_e$, and this is irrespective of the type of a transition, i.e. allowed, forbidden or inter-combination. This is further confirmed by their figs. 6 and 7, and note particularly the sharp fall (up to a factor of 5) in $\Upsilon$ values for the 1--11 (3s$^2$3p~$^2$P$^o_{1/2}$ -- 3s$^2$3d~$^2$D$_{3/2}$)  allowed transition in the left panel of fig.~6 for Ga~XIX, As~XXI and Br~XXIII --  see also present Figs. 1e--5e. This is not a normal behaviour as the $\Omega$ values for allowed transitions are known to increase with increasing energy ($\Omega$ $\sim$ (4$\omega_i$f/$\Delta$E$_{ij}$)$\times$log (E), where $\omega_i$ is the statistical weight of the lower level $i$, f and $\Delta$E$_{ij}$ are the oscillator strength and transition  energy, respectively, and E is the incident electron energy), and subsequently therefore the $\Upsilon$ values too  increase with increasing T$_e$.  Interestingly,  Wang and Jiang \cite{wj} have themselves recognised this fact by stating that ``For dipole-allowed transitions, the collision strength diverges logarithmically as the energy tends to infinity; for non- dipole-allowed transitions, it tends to a constant and for forbidden transitions, the collision strength shows a decreasing tendency in the infinite energy limit.", and yet they have not followed these trends for the $\Upsilon$ values, although they have shown increasing values of $\Omega$, in a limited energy range below 140~Ryd, for some allowed transitions in their figs.~3 and 5.

The reason for decreasing values of $\Upsilon$ of  Wang and Jiang \cite{wj}   with increasing T$_e$ is not difficult to understand. This is because they have calculated values of $\Omega$ up to 4 times the ionization energy of an ion, i.e. up to a maximum of 280~Ryd, or more specifically up to about 200, 220, 240, 260, and 280~Ryd for Ga~XIX, Ge~XX, As~XXI, Se~XXII, and Br~XXII, respectively. These limited energies are {\em not} sufficient to calculate $\Upsilon$ values up to a temperature of 10$^8$~K, which is equivalent to about 633.37~Ryd. Since neither have they calculated $\Omega$ values up to higher enough energies (say up to about 2000~Ryd) nor have they extrapolated their data to higher energies, their $\Upsilon$ values sharply decrease at higher temperatures, and hence are {\em incorrect}.

\section{Energy levels and radiative rates}
\label{calc}

For the calculations of wavefunctions, and subsequently the energy levels and A-values, Wang and Jiang \cite{wj} have included expansion among 15 configurations, namely 3s$^2$3p, 3s3p$^2$, 3s$^2$3d, 3s3p$^2$, 3s3d$^2$, 3p$^3$, 3s3p3d, 3s$^2$4$\ell$, and 3s3p4$\ell$. An `extended average level' (EAL) approximation has been adopted in which the weighted (proportional to 2$j$+1) trace of the Hamiltonian matrix is minimized. This approximation generally gives a fairly accurate estimate of the energies for most levels, and is confirmed by the comparisons shown for a few levels in their table~A,  between their calculated results and the compilations of experimental energies, available on the NIST (National Institute of Standards and Technology)  website {\tt http://www.nist.gov/pml/data/asd.cfm}. Furthermore, there are no appreciable discrepancies with the other available theoretical results -- see their tables~1--5, in which energies are listed for the lowest 82 levels although the above noted 15 configurations generate 163 in total.

In our calculations with the Flexible Atomic Code (FAC) of Gu \cite{fac}, available on the website {\tt https://www-amdis.iaea.org/FAC/}, we have included the same 15 configurations as by Wang and Jiang \cite{wj}. Since their A-values are restricted to the lowest 40 levels, and those of $\Upsilon$ to only 30, in Table~A we compare our calculated energies with their GRASP calculations for the lowest 30 levels alone, but for all 5 ions, namely Ga~XIX, Ge~XX, As~XXI, Se~XXII, and Br~XXII. This comparison confirms that there are no appreciable discrepancies between the two sets of calculated energies, although there are a few minor differences in some (closely lying) level orderings -- see for example, levels 20/21 and 26/27 of Ga~XIX. However, these minor differences will not affect our subsequent calculations,  comparisons or conclusions, because our focus is mainly on the $\Upsilon$ values.

Since $\Omega$ values for allowed transitions generally increase with increasing energy, are mostly dependent on oscillator strengths and transition energies, as already stated, often have larger magnitudes, and are not normally dominated in the thresholds region by  resonances (if any), it is much easier to compare this parameter (and subsequently $\Upsilon$) between any two calculations. For this reason, in Table~B we list the A-Values of Wang and Jiang \cite{wj} for 5 allowed transitions from the lowest two levels of the ground configuration 3s$^2$3p of Ga~XIX, Ge~XX, As~XXI, Se~XXII, and Br~XXII. To be specific these transitions are: 1--6 (3s$^2$3p~$^2$P$^o_{1/2}$ -- 3s3p$^2$~$^2$D$_{3/2}$), 1--8 (3s$^2$3p~$^2$P$^o_{1/2}$ -- 3s3p$^2$~$^2$P$_{1/2}$), 1--9 (3s$^2$3p~$^2$P$^o_{1/2}$ -- 3s3p$^2$~$^2$S$_{1/2}$), 1--10 (3s$^2$3p~$^2$P$^o_{1/2}$ -- 3s3p$^2$~$^2$P$_{3/2}$), 1--11 (3s$^2$3p~$^2$P$^o_{1/2}$ -- 3s$^2$3d~$^2$D$_{3/2}$), 2--6 (3s$^2$3p~$^2$P$^o_{3/2}$ -- 3s3p$^2$~$^2$D$_{3/2}$), 2--7 (3s$^2$3p~$^2$P$^o_{3/2}$ -- 3s3p$^2$~$^2$D$_{5/2}$), 2--9 (3s$^2$3p~$^2$P$^o_{3/2}$ -- 3s3p$^2$~$^2$S$_{1/2}$), 2--10 (3s$^2$3p~$^2$P$^o_{3/2}$ -- 3s3p$^2$~$^2$P$_{3/2}$), and 2--11 (3s$^2$3p~$^2$P$^o_{3/2}$ -- 3s$^2$3d~$^2$D$_{3/2}$). Our corresponding results with FAC for both A- and f- values are also listed in this table for a ready comparison. For all these transitions, and many more not listed here, there are no appreciable discrepancies between the two calculations. As a result of this, the subsequent values of $\Omega$ and $\Upsilon$ should also be comparable at all energies and temperatures, respectively.

\section{Collision strengths and effective collision strengths}

The $\Omega$ values are often  averaged over a  {\em Maxwellian} distribution of electron velocities to determine the corresponding $\Upsilon$ values -- see eq. (1) of \cite{wj}. In Figs.~1--5 (a and b) we show the $\Upsilon$ values of Wang and Jiang \cite{wj} for the 10 transitions, listed in Table~B, for all 5 Al-like ions, Ga~XIX, Ge~XX, As~XXI, Se~XXII, and Br~XXII. For all transitions their $\Upsilon$ values invariably decrease (that too sharply by up to a factor of 5)  particularly   at T$_e$ $>$ 10$^7$~K. Additionally, for some transitions, such as 1--11 of As~XXI and  Se~XXII, their $\Upsilon$ values show an anomalous behaviour at some temperatures, as shown in Figs.~3a and 4a. This is because they have prepared their tables by hand and as a result the $\Upsilon$ values have been quoted lower by an order of magnitude. Similar anomalies were also detected in their earlier work on N-like ions \cite{nlike} -- see \cite{km1} for details. As a result of this their reported data for $\Upsilon$ values cannot be confidently used in any application.

In Figs.~1--5 (c and d) we show our results of $\Omega$ with the FAC code, which is fully relativistic and calculates collisional data with the {\em distorted-wave} (DW) method. It has been demonstrated in several of our earlier papers (see \cite{km2} and \cite{km3} for examples and references) that the results for $\Omega$ are often comparable with those with DARC for most of the transitions, particularly at energies above thresholds. As expected, values of $\Omega$ vary smoothly and increase with increasing energy, for all allowed transitions and for all ions. Subsequently, we expect a similar behaviour of $\Upsilon$ values, which is further confirmed in Figs.~1--5 (e and f) for the 10 transitions of all 5 Al-like ions under discussion. By the comparisons shown in these figures, it is estimated that the $\Upsilon$ values reported by  Wang and Jiang \cite{wj} are underestimated by up to (almost) an order of magnitude, particularly towards the higher end of the temperature range. Furthermore, there are a few more anomalies in the $\Upsilon$ data of \cite{wj}. For example, as seen in their tables~6--10, A- (and hence the f-) values for the 2--7 transitions are higher than for 2--8, by up to three orders of magnitude (depending on the ion), but their $\Upsilon$ values are in the reverse orders, as seen in their tables~11--15. This is in spite of the fact that energies for these transitions are comparable within about 10\%. For this reason, the $\Upsilon$ values of Wang and Jiang \cite{wj}  shown in Figs.~1f--5f for the 2--7 transitions are in fact for 2--8.

\section{Conclusions}

In this paper we have compared our atomic data for energy levels, radiative rates and effective collision strengths with those of  Wang and Jiang \cite{wj}, for 5 Al-like ions, namely  Ga~XIX, Ge~XX, As~XXI, Se~XXII, and Br~XXII. There are mainly two differences between the present and the earlier calculations. Firstly, they have adopted the GRASP  code for the determinations of energy levels and A-values, and subsequently DARC for the calculations of $\Omega$ and $\Upsilon$. On the other hand, we have adopted the FAC code which determines all parameters, but in a similar relativistic manner. However, the differences in the two codes do not generally affect the calculations of these parameters, for most levels and/or transitions. This is also confirmed by the comparisons shown in Tables~A and B. Conversely, it can be confidently said that there are no significant anomalies in the energy levels and A-values of \cite{wj}. The other difference is in the determination of $\Upsilon$ values, because they have resolved resonances in $\Omega$ in the thresholds energy regions, whereas we have not. However, their contributions are mostly noticeable for forbidden (and some inter-combination) transitions, {\em and} at lower temperatures. At high temperatures ($\ge$ 10$^6$~K) where the data for $\Upsilon$ are mostly required for these ions, particularly for applications in fusion plasmas, the contributions of resonances are (almost) negligible. Therefore, results from the two different codes should be comparable for $\Upsilon$ values, especially at high(er) temperatures. Unfortunately, this is not the case. Values of $\Upsilon$ reported by  Wang and Jiang \cite{wj} are underestimated by up to an order of magnitude! This is mainly because they have calculated values of $\Omega$ in a limited energy range (up to 280~Ryd, depending on the ion), but have determined $\Upsilon$ up to about 630~Ryd (10$^8$~K). As a results of this, their $\Upsilon$ values decrease for {\em all types of} transitions at higher temperatures. 

Apart from the major errors in the $\Upsilon$ values of  Wang and Jiang \cite{wj}, there are a few more anomalies. For some transitions and at some temperatures their $\Upsilon$ values are incorrectly listed, and are lower by an order of magnitude. This is because they have prepared the tables by hand rather than generating through computer. Similarly, for some transitions (such as 2--7 and 2--8) their A- and $\Upsilon$ values are incompatible, and are in reverse order. Furthermore, they have performed calculations among 163 levels of 15 configurations, but have reported only limited data, mostly for transitions  from the lowest two levels to up to 30. This hampers applications in modelling of plasmas, because a complete set of data for all transitions are preferred and useful.  Since the reported data of Wang and Jiang \cite{wj} are anyway erroneous, it is advised to use $\Upsilon$ data from FAC till better calculations can be performed.  The FAC code is fairly reliable, is highly efficient, is easy to use, and is freely available. Therefore, potential users may either generate data themselves or can obtain from the first author on request.

Calculations of collisional data with DARC, or other versions of the R-matrix code,  are a preferred choice because of the inclusion of resonances in the determination of $\Upsilon$ values. However, such calculations are time consuming, require large computational resources, and more importantly, require expertise, training or supervision, because a number of checks at several stages are desired to assess the accuracy of the data. In the absence of that chances of error/s remain as has been highlighted and explained in several of our earlier papers, see in particular \cite{km2},\cite{km3}. We hope more accurate collisional data for Al-like ions will be available in the future.

\ack
We thank Dr. Robert Ryans  for his help in providing computational facilities.

\begin{figure*}
\includegraphics[angle=-90,width=0.9\textwidth]{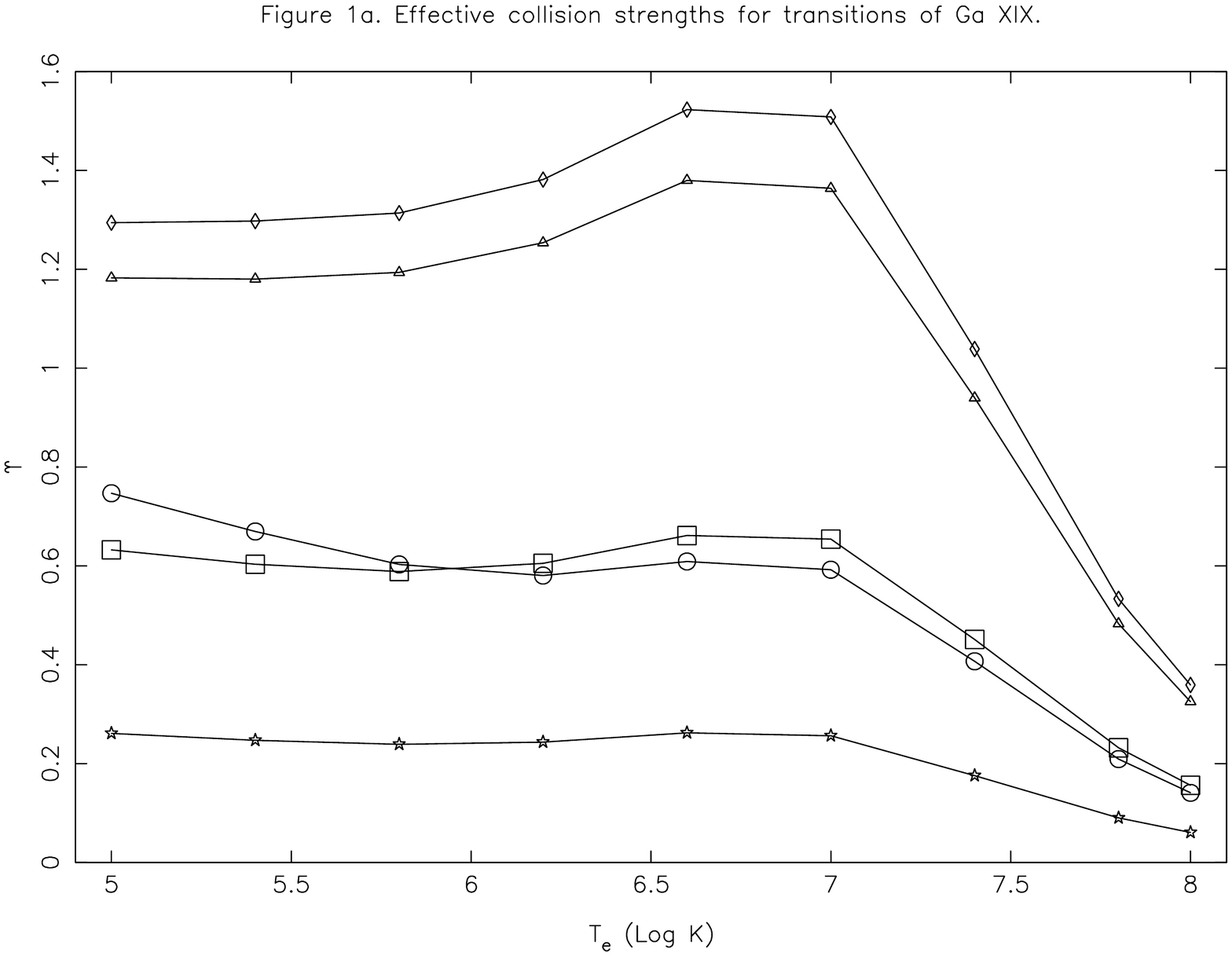}
 \vspace{-1.5cm}
 \caption{}
 \end{figure*}

\setcounter{figure}{0}
 \begin{figure*}
\includegraphics[angle=-90,width=0.9\textwidth]{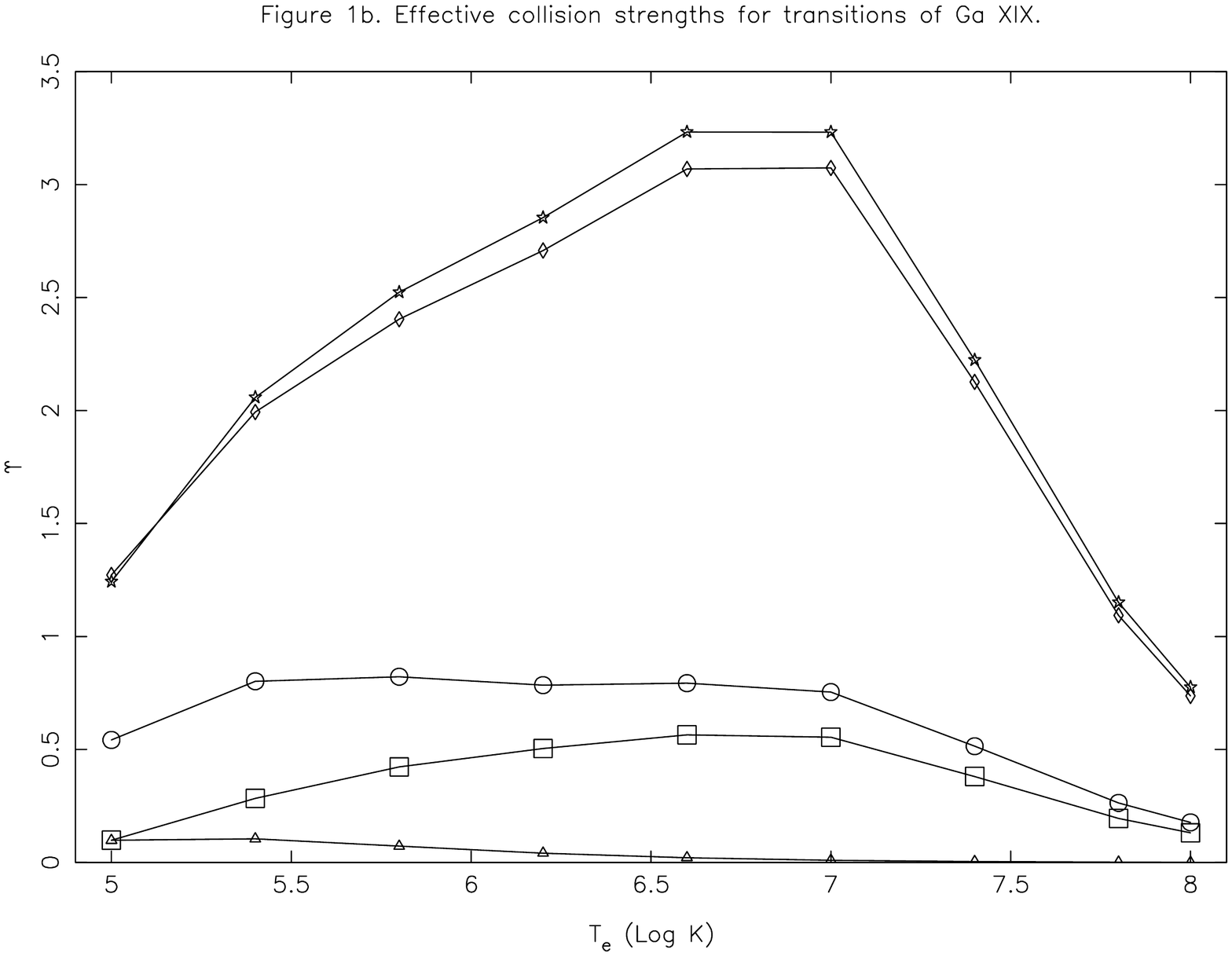}
 \vspace{-1.5cm}
\caption{}
 \end{figure*}

\setcounter{figure}{0}
 \begin{figure*}
\includegraphics[angle=-90,width=0.9\textwidth]{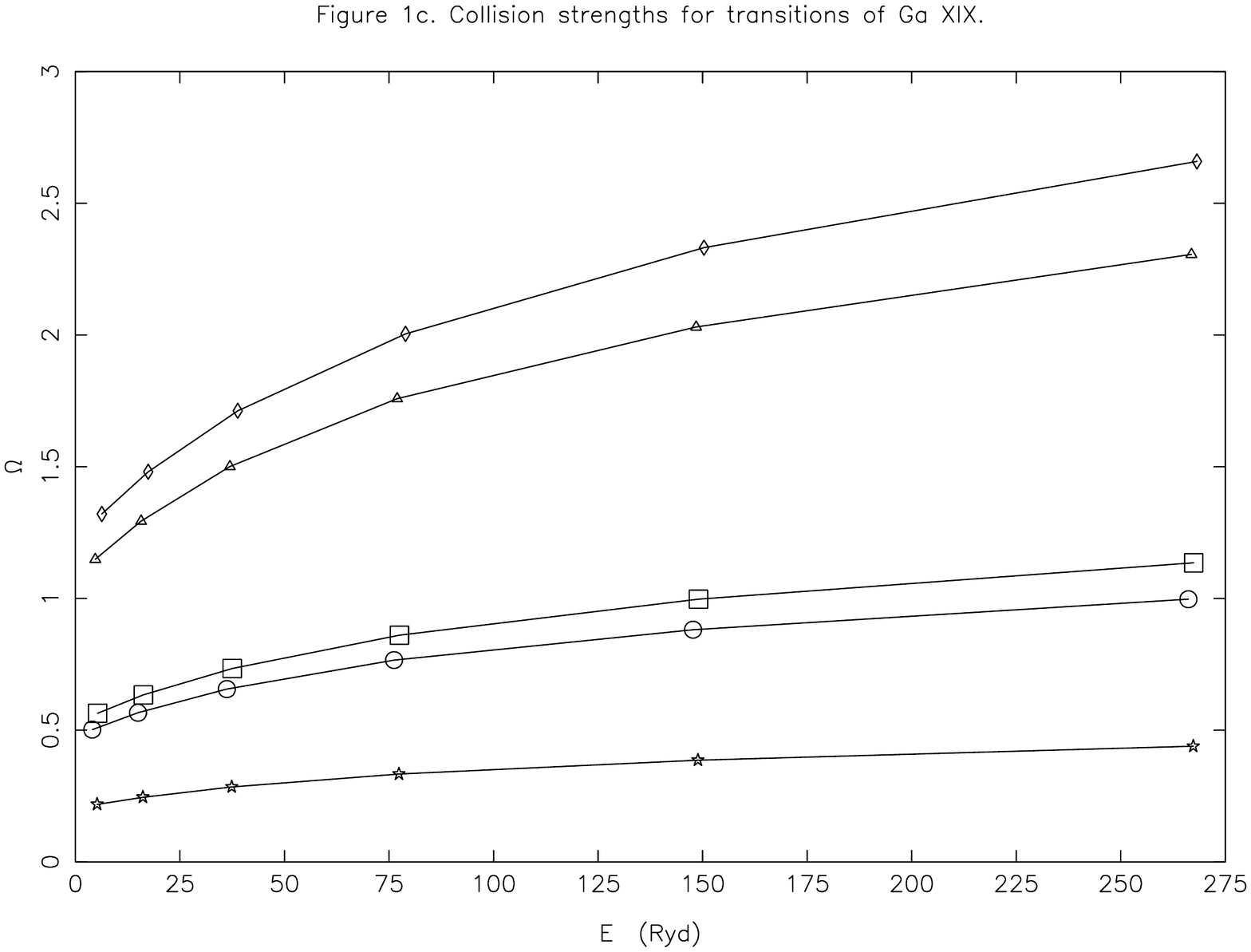}
 \vspace{-1.5cm}
\caption{}
 \end{figure*}

\setcounter{figure}{0}
 \begin{figure*}
\includegraphics[angle=-90,width=0.9\textwidth]{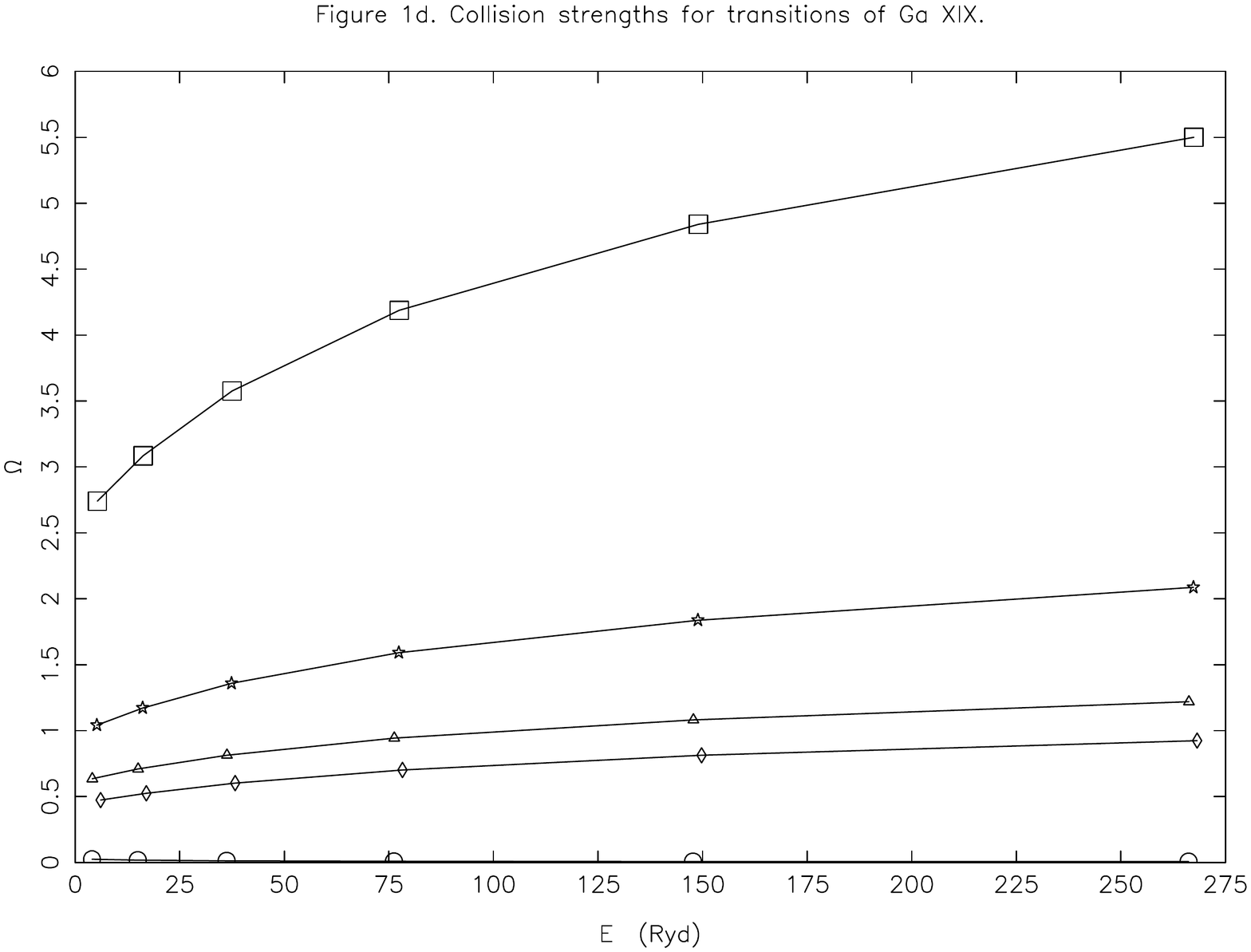}
 \vspace{-1.5cm}
\caption{}
 \end{figure*}

\setcounter{figure}{0}
 \begin{figure*}
\includegraphics[angle=-90,width=0.8\textwidth]{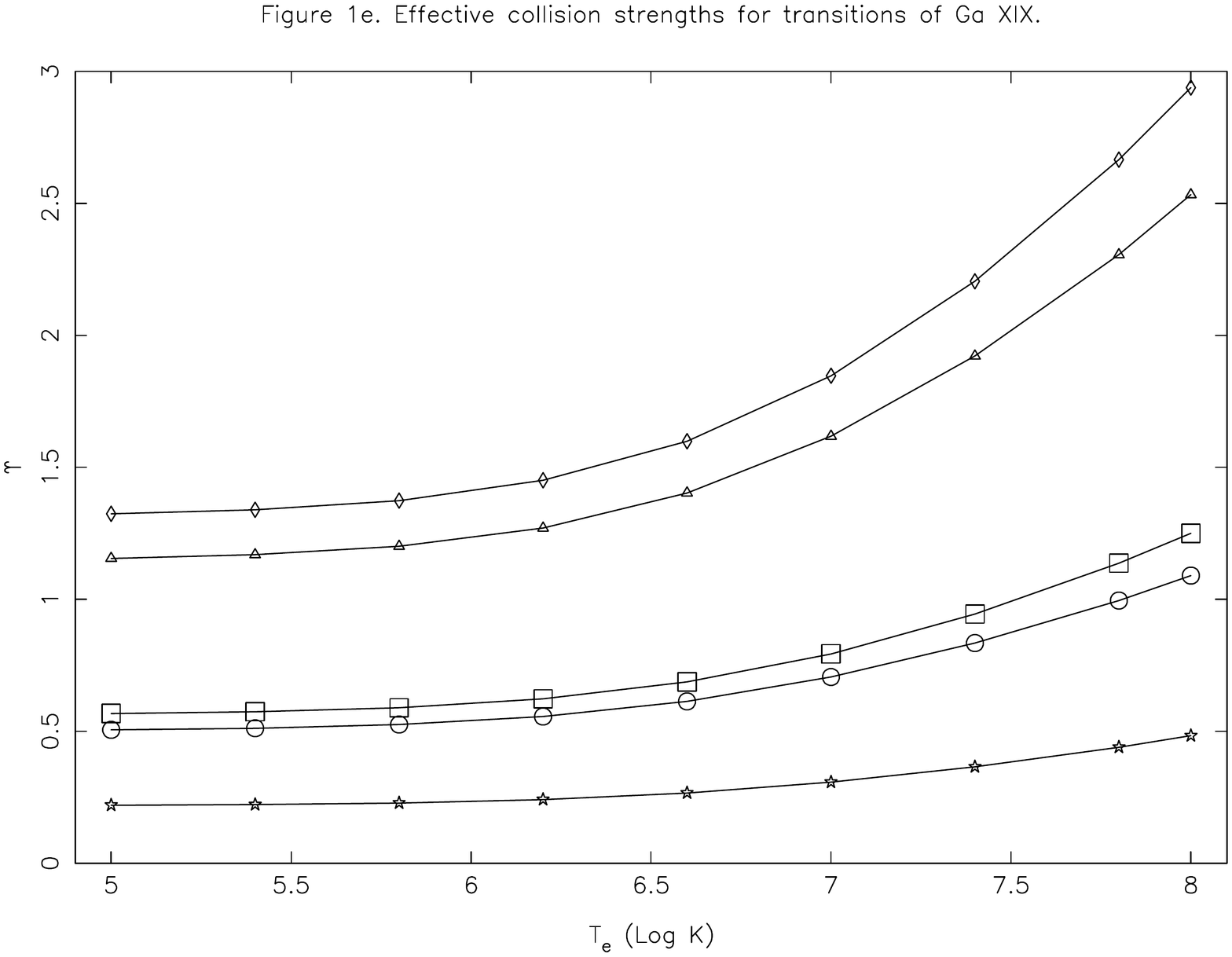}
 \vspace{-1.5cm}
\caption{}
 \end{figure*}

\setcounter{figure}{0}
 \begin{figure*}
\includegraphics[angle=-90,width=0.8\textwidth]{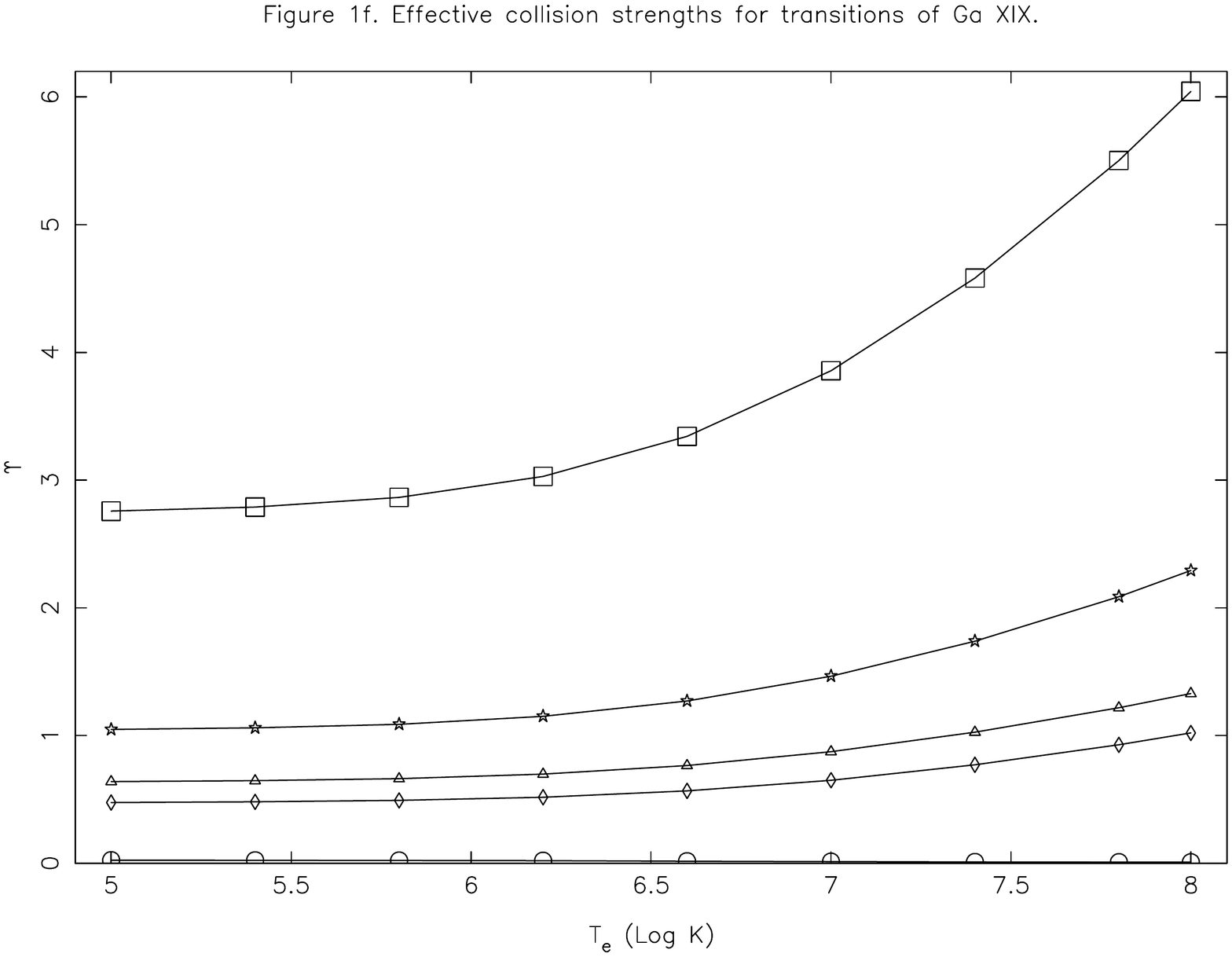}
 \vspace{-1.5cm}
\caption{Effective collision strengths of \cite{wj} (a):  from the ground level (1: 3s$^2$3p~$^2$P$^o_{1/2}$) to the excited levels (6; circles: 3s3p$^2$~$^2$D$_{3/2}$,   
8; triangles: 3s3p$^2$~$^2$P$_{1/2}$,  9; stars: 3s3p$^2$~$^2$S$_{1/2}$,  10; squares: 3s3p$^2$~$^2$P$_{3/2}$,  and 11; diamonds: 3s$^2$3d~$^2$D$_{3/2}$) of Ga~XIX. Corresponding data in (b) are from the level 2 (3s$^2$3p~$^2$P$^o_{3/2}$). In (c) and (d) are our collision strengths with FAC and in (e) and (f) the effective collision strengths for the same corresponding transitions. }
 \end{figure*}

\setcounter{figure}{1}
\begin{figure*}
\includegraphics[angle=-90,width=0.8\textwidth]{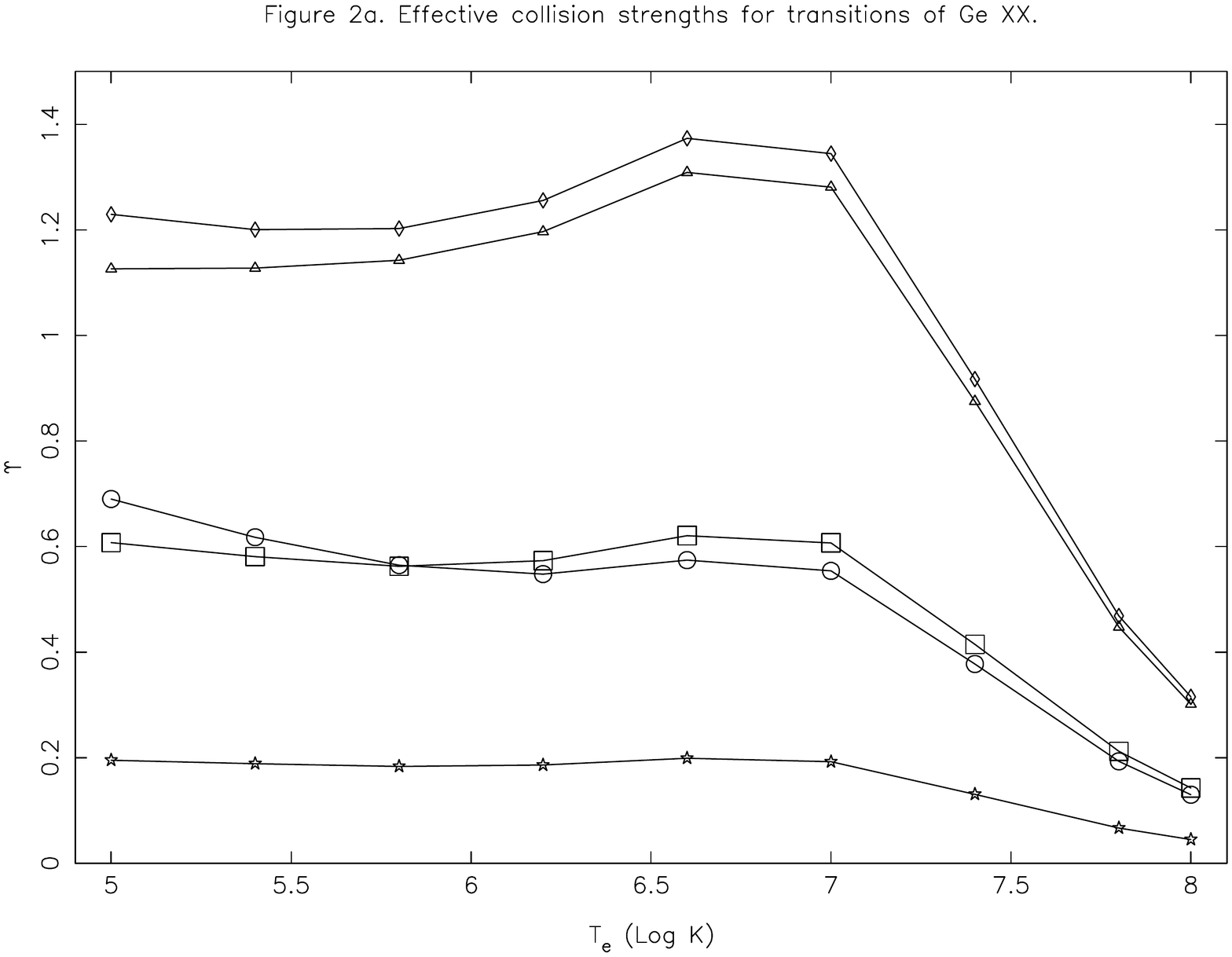}
 \vspace{-1.5cm}
 \caption{}
 \end{figure*}

\setcounter{figure}{1}
 \begin{figure*}
\includegraphics[angle=-90,width=0.8\textwidth]{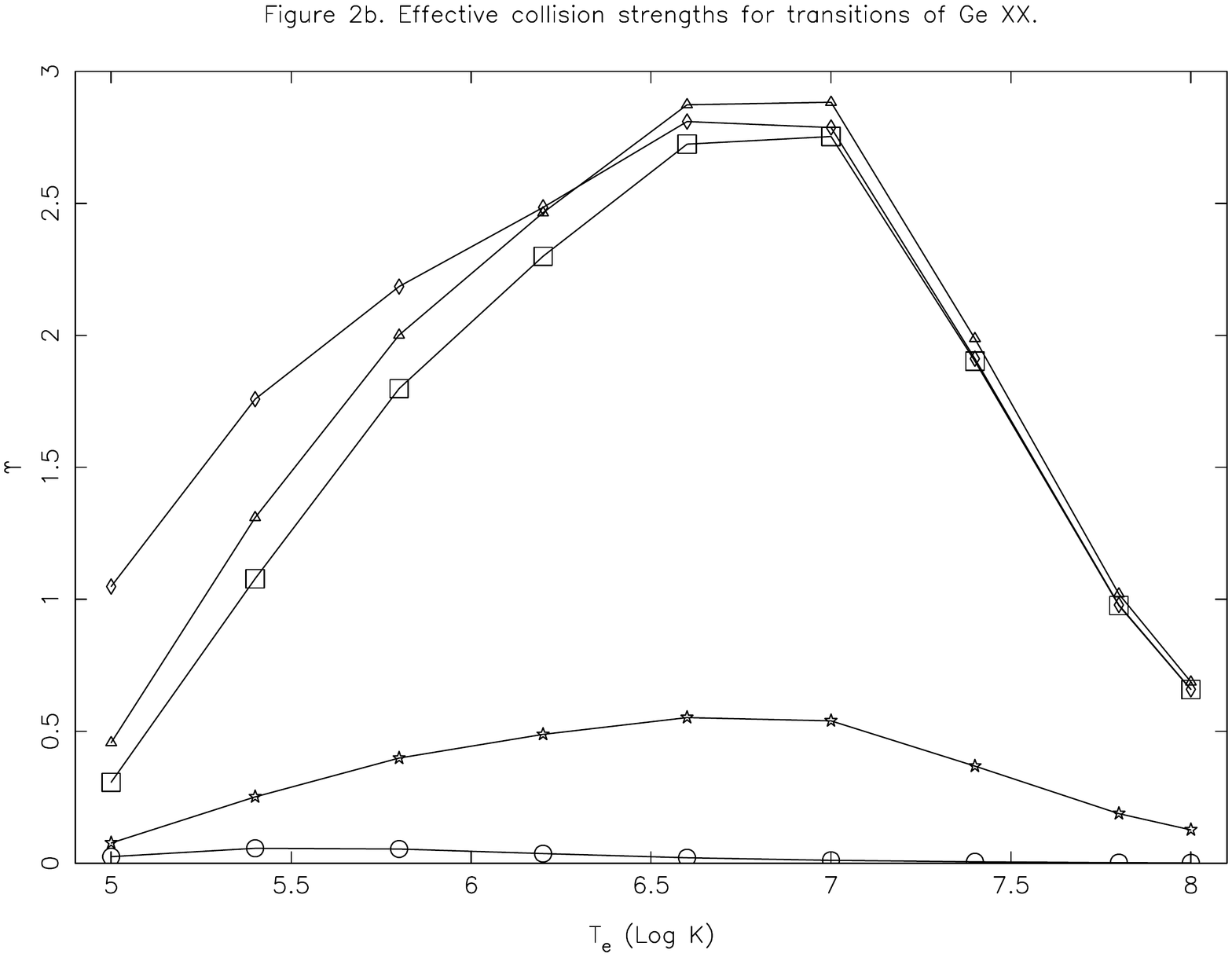}
 \vspace{-1.5cm}
\caption{}
 \end{figure*}

\setcounter{figure}{1}

 \begin{figure*}
\includegraphics[angle=-90,width=0.8\textwidth]{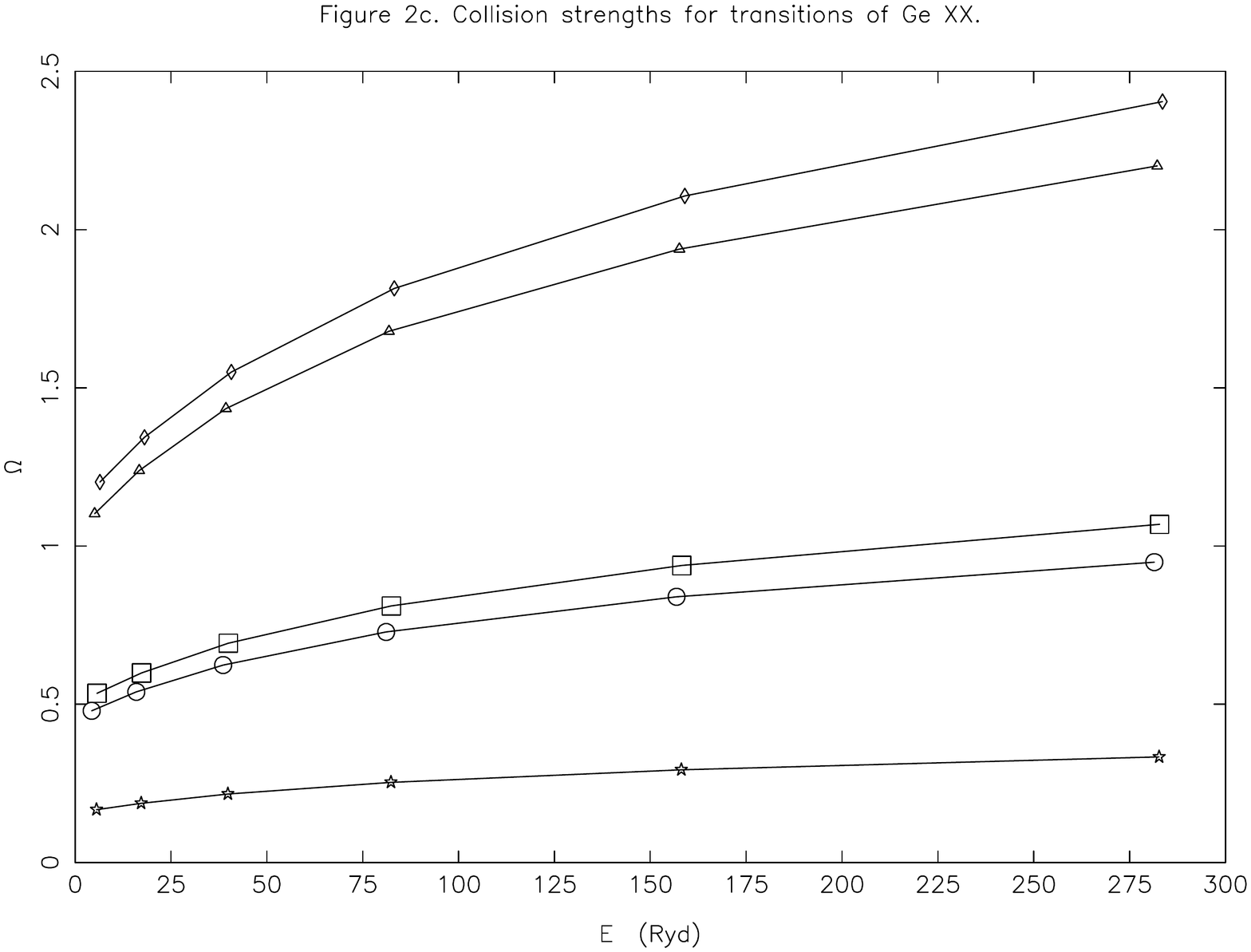}
 \vspace{-1.5cm}
\caption{}
 \end{figure*}

\setcounter{figure}{1}
 \begin{figure*}
\includegraphics[angle=-90,width=0.8\textwidth]{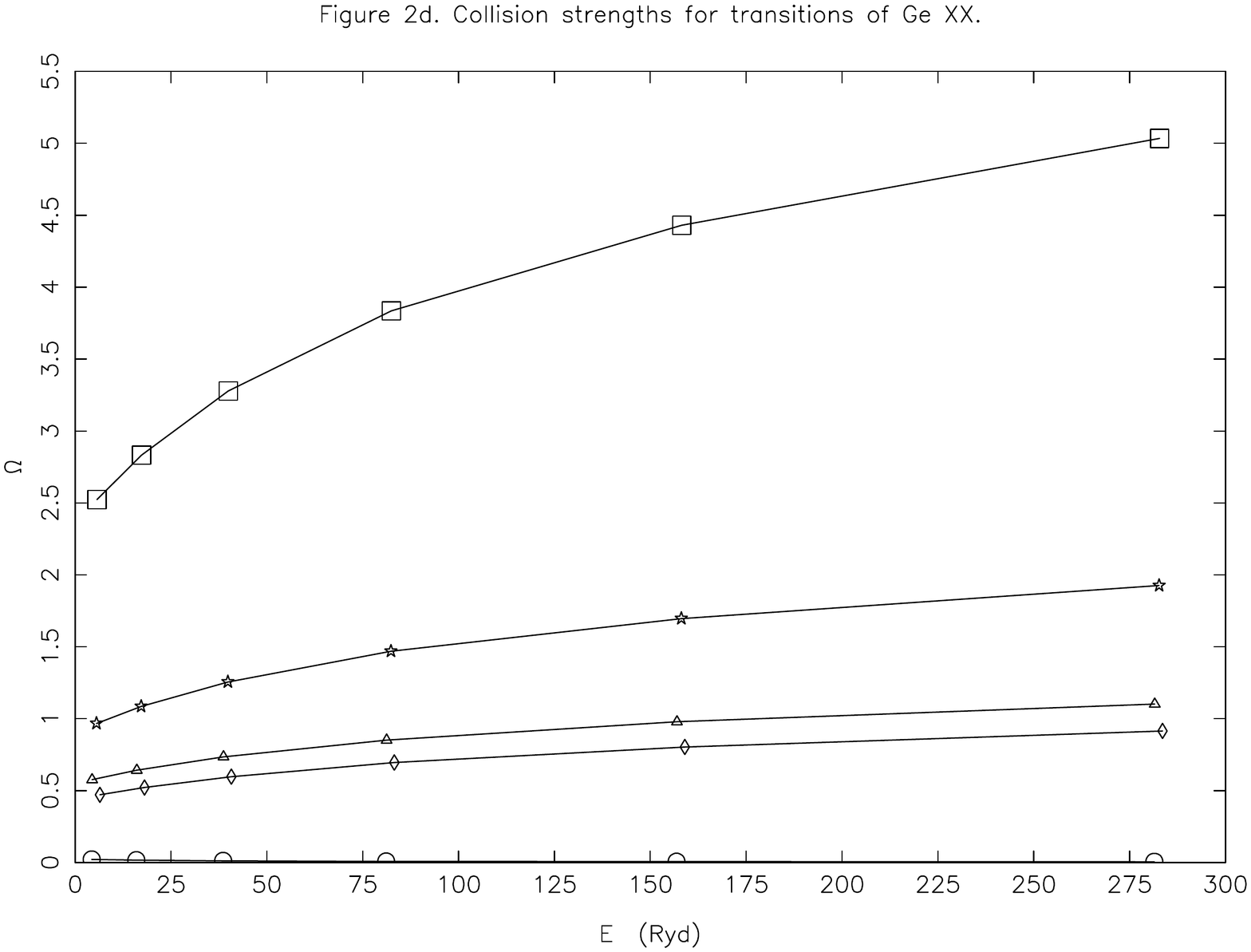}
 \vspace{-1.5cm}
\caption{Collision strengths for the 1--3 (}
 \end{figure*}

\setcounter{figure}{1}
 \begin{figure*}
\includegraphics[angle=-90,width=0.8\textwidth]{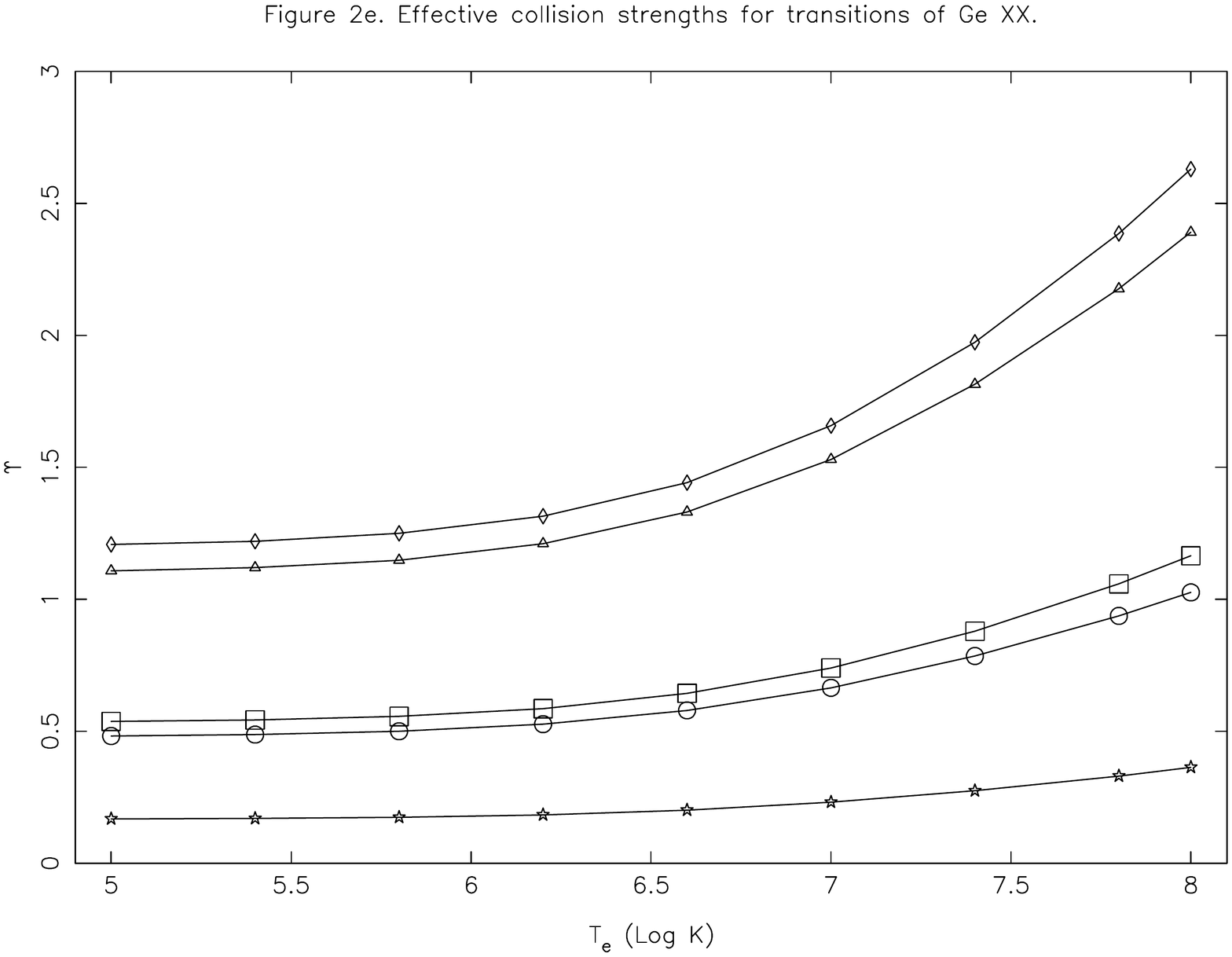}
 \vspace{-1.5cm}
\caption{}
 \end{figure*}

\setcounter{figure}{1}
 \begin{figure*}
\includegraphics[angle=-90,width=0.8\textwidth]{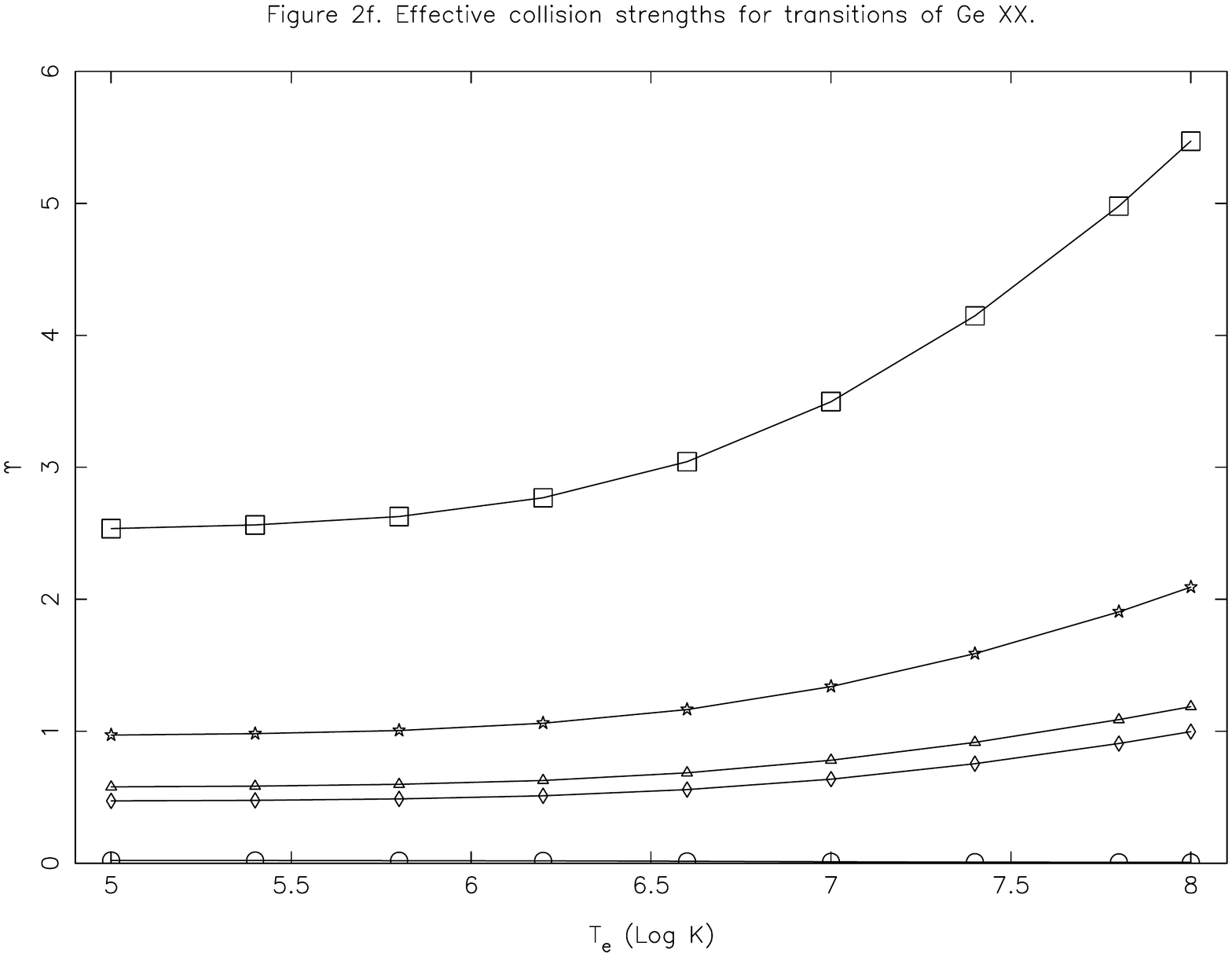}
 \vspace{-1.5cm}
\caption{Effective collision strengths of \cite{wj} (a):  from the ground level (1: 3s$^2$3p~$^2$P$^o_{1/2}$) to the excited levels (6; circles: 3s3p$^2$~$^2$D$_{3/2}$,   
8; triangles: 3s3p$^2$~$^2$P$_{1/2}$,  9; stars: 3s3p$^2$~$^2$S$_{1/2}$,  10; squares: 3s3p$^2$~$^2$P$_{3/2}$,  and 11; diamonds: 3s$^2$3d~$^2$D$_{3/2}$) of Ge~XX. Corresponding data in (b) are from the level 2 (3s$^2$3p~$^2$P$^o_{3/2}$). In (c) and (d) are our collision strengths with FAC and in (e) and (f) the effective collision strengths for the same corresponding transitions.}
 \end{figure*}

 \setcounter{figure}{2}
\begin{figure*}
\includegraphics[angle=-90,width=0.8\textwidth]{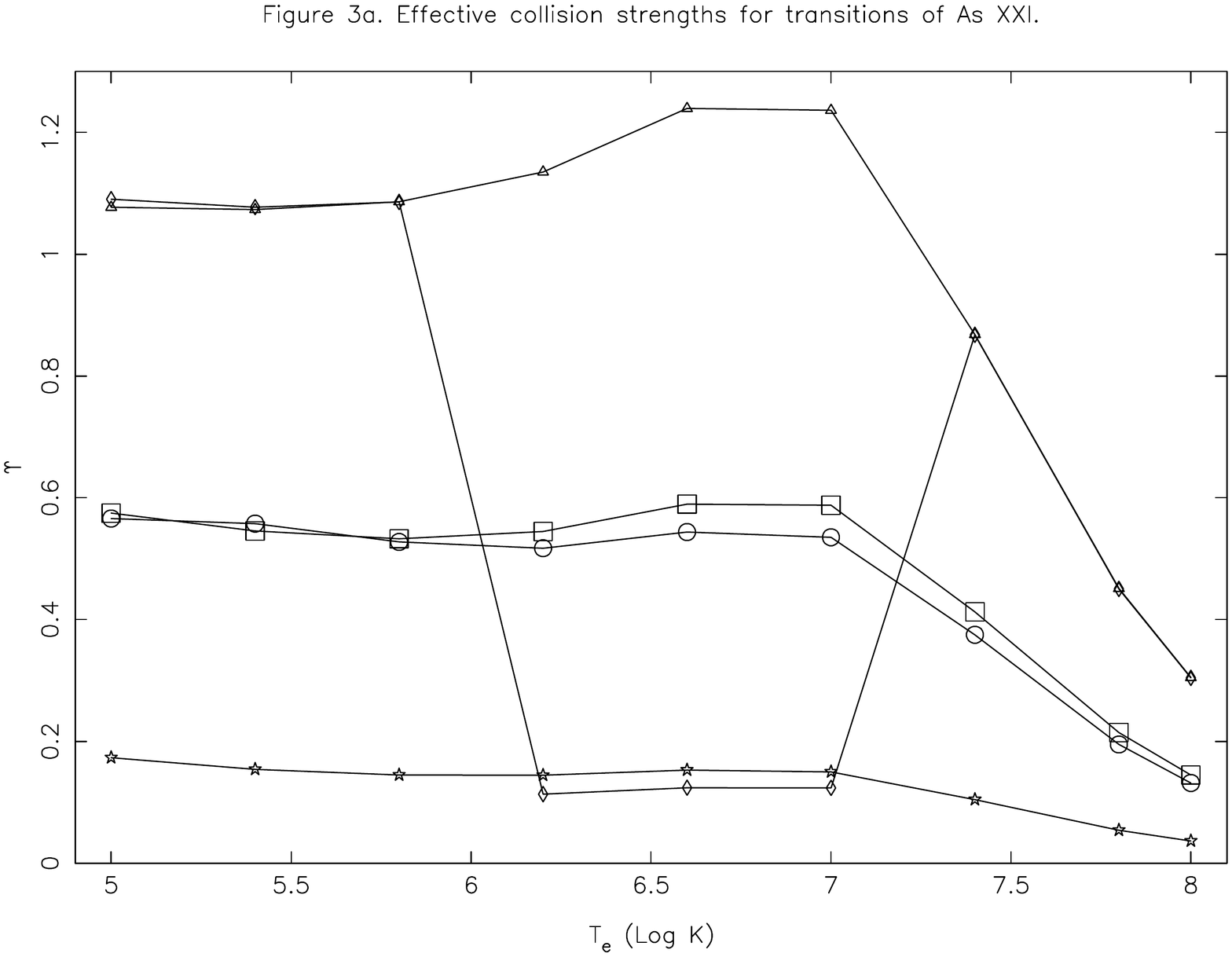}
 \vspace{-1.5cm}
 \caption{}
 \end{figure*}

\setcounter{figure}{2}
 \begin{figure*}
\includegraphics[angle=-90,width=0.8\textwidth]{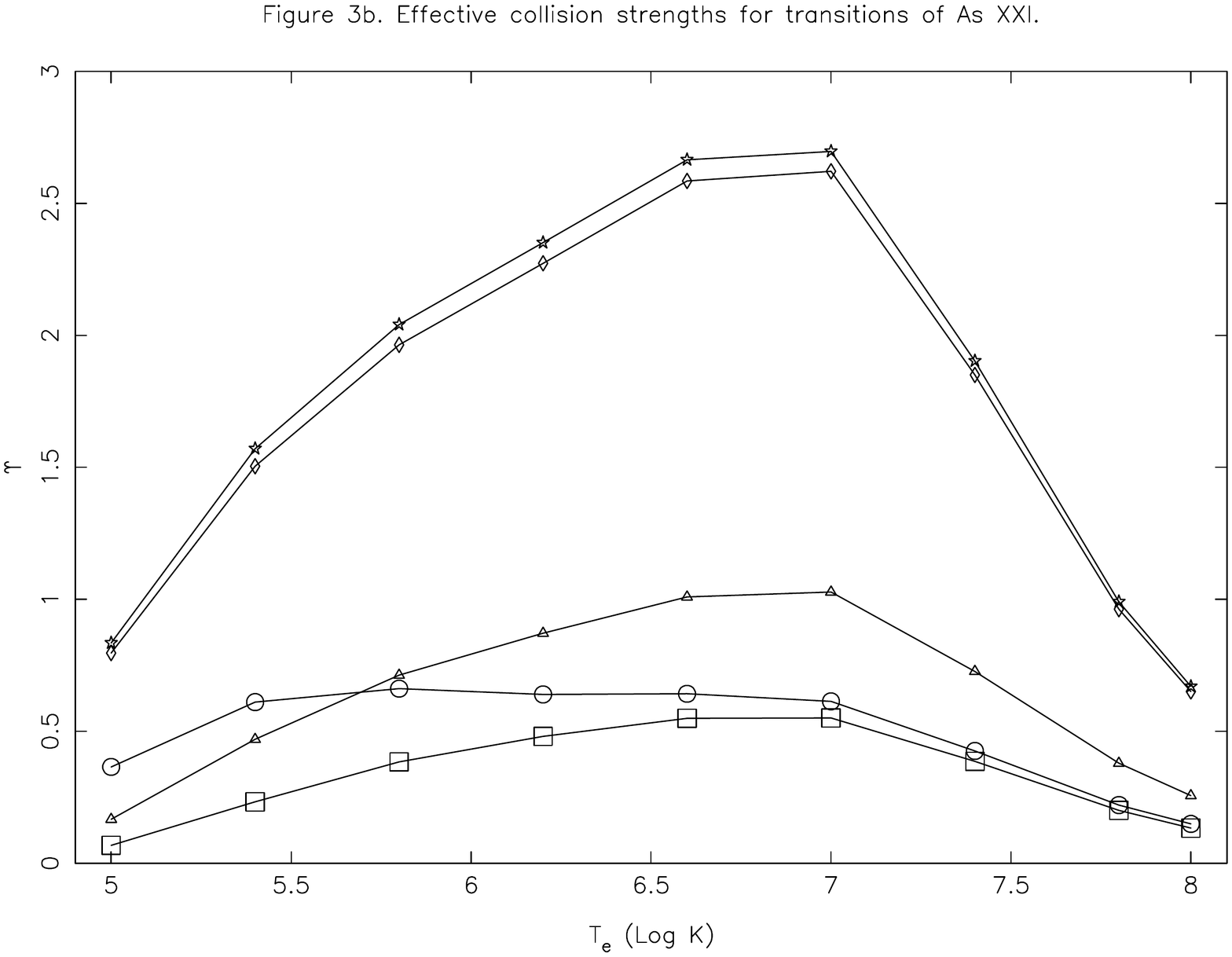}
 \vspace{-1.5cm}
\caption{}
 \end{figure*}

\setcounter{figure}{2}

 \begin{figure*}
\includegraphics[angle=-90,width=0.8\textwidth]{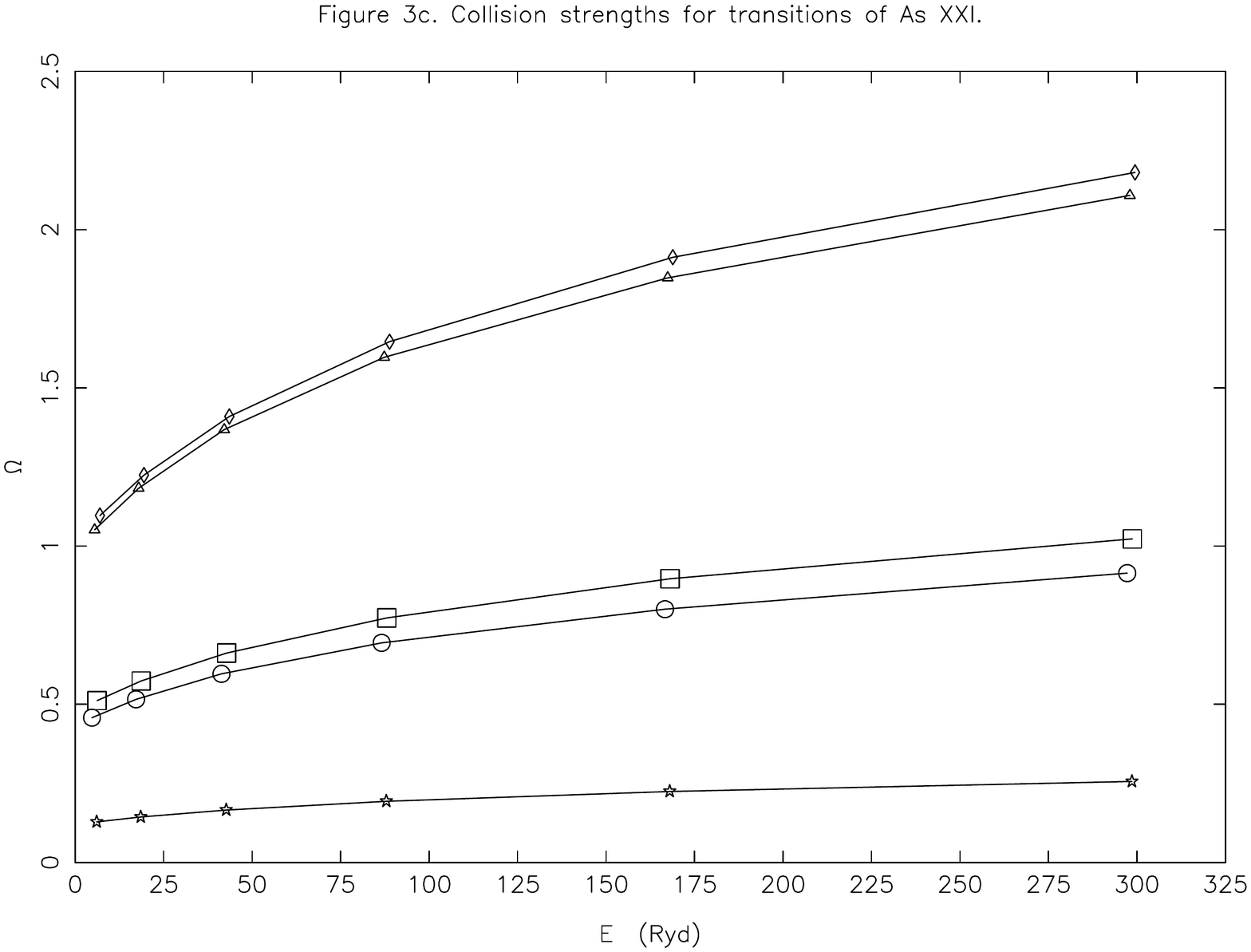}
 \vspace{-1.5cm}
\caption{}
 \end{figure*}

\setcounter{figure}{2}
 \begin{figure*}
\includegraphics[angle=-90,width=0.8\textwidth]{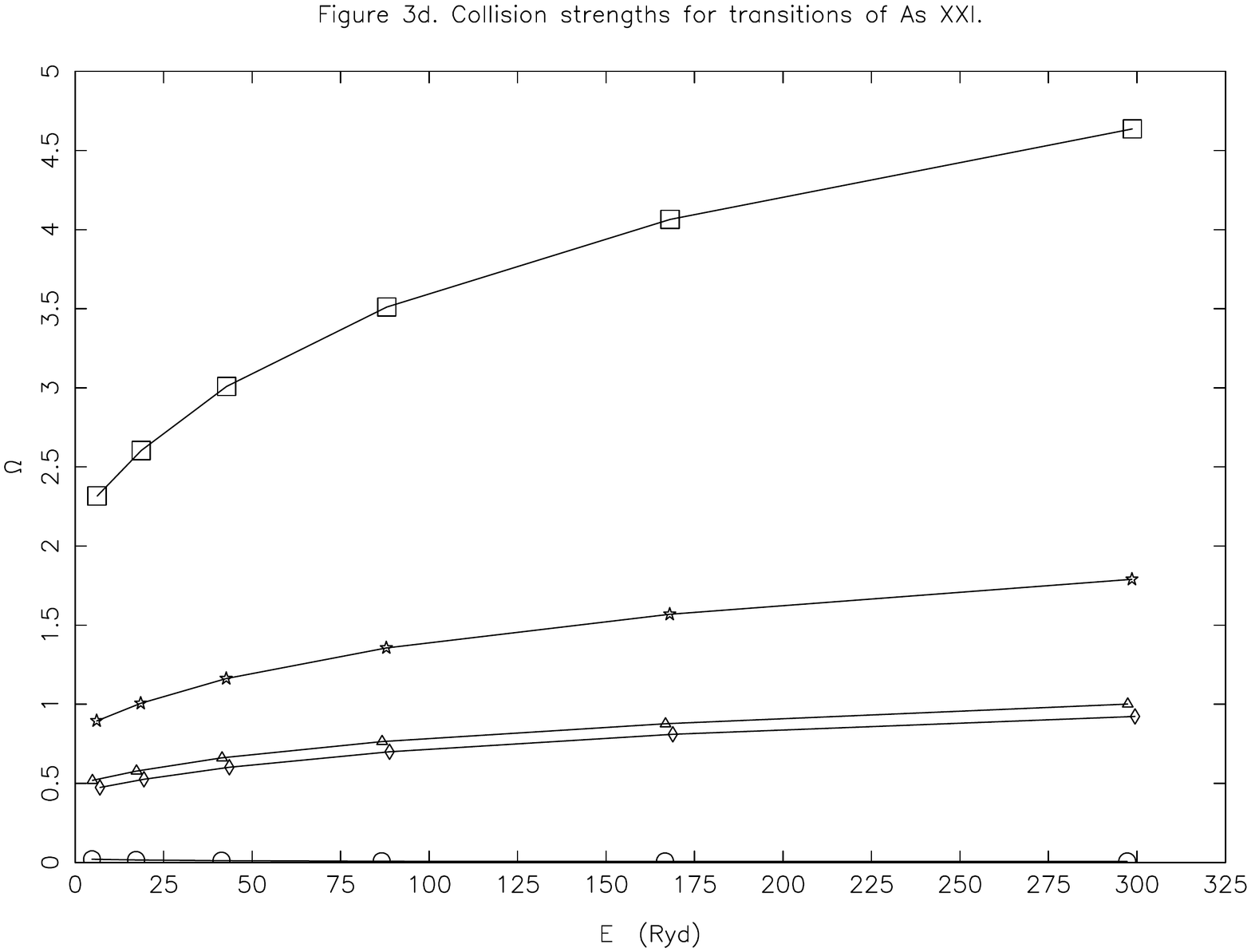}
 \vspace{-1.5cm}
\caption{}
 \end{figure*}

\setcounter{figure}{2}
 \begin{figure*}
\includegraphics[angle=-90,width=0.8\textwidth]{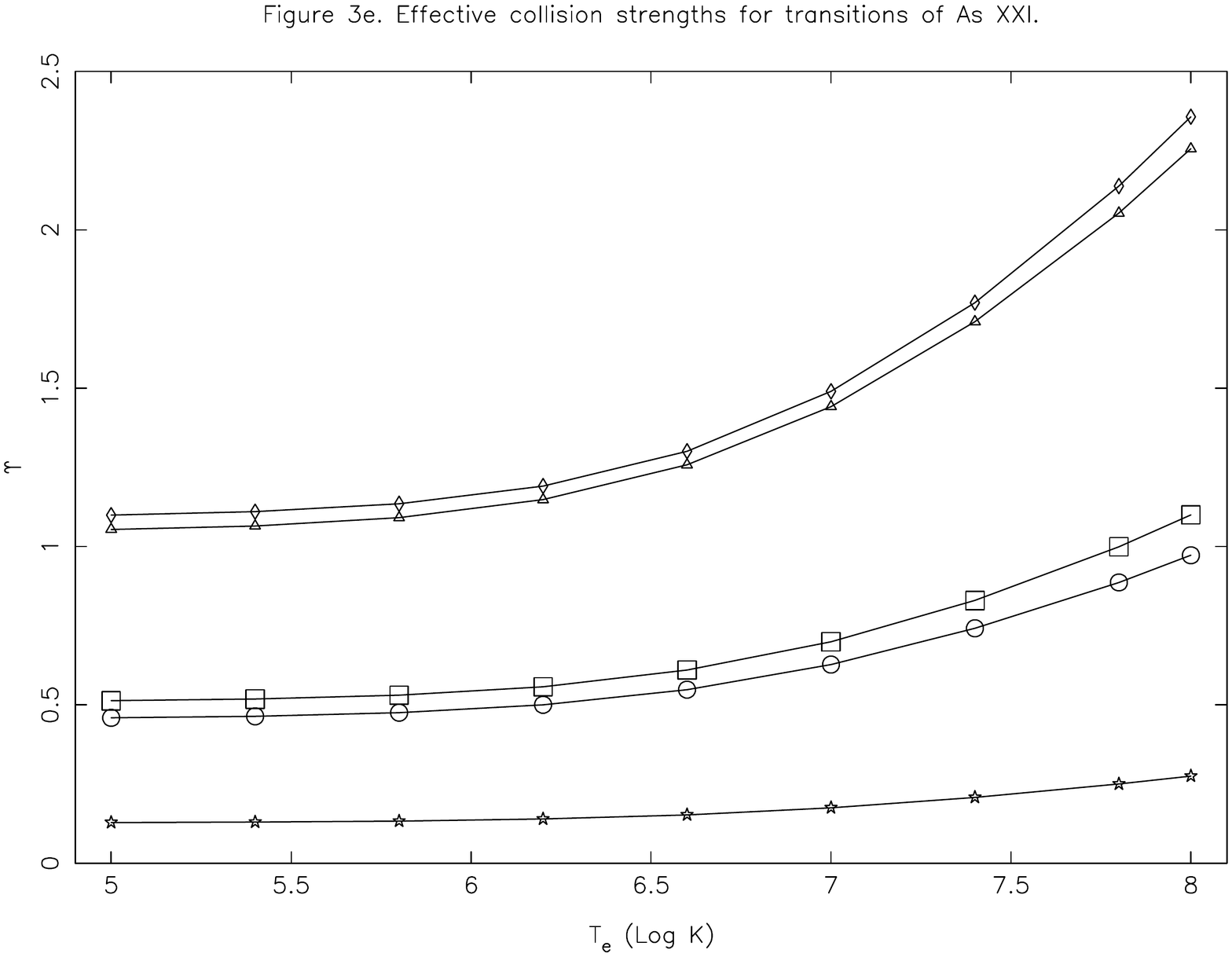}
 \vspace{-1.5cm}
\caption{}
 \end{figure*}

\setcounter{figure}{2}
 \begin{figure*}
\includegraphics[angle=-90,width=0.8\textwidth]{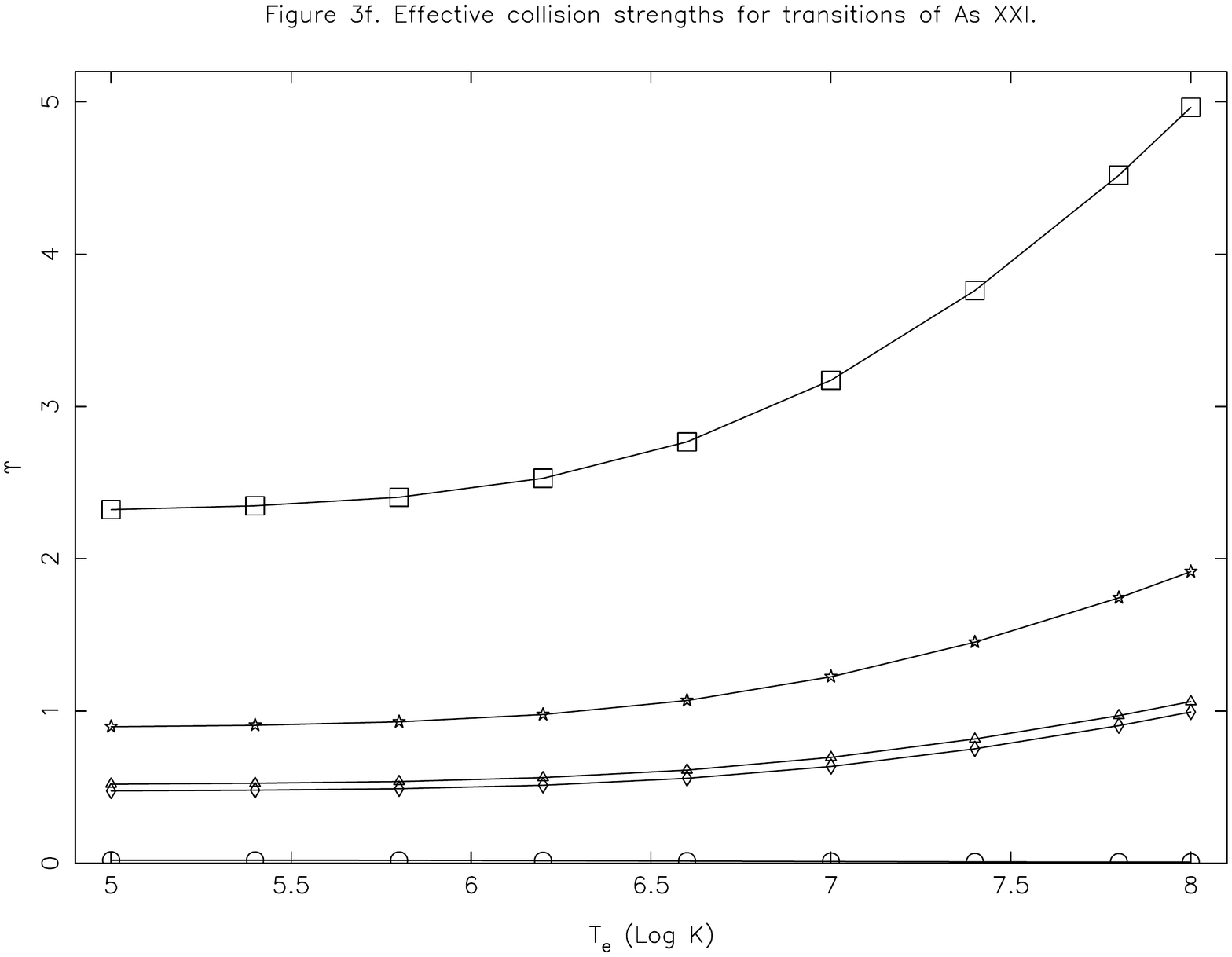}
 \vspace{-1.5cm}
\caption{Effective collision strengths of \cite{wj} (a):  from the ground level (1: 3s$^2$3p~$^2$P$^o_{1/2}$) to the excited levels (6; circles: 3s3p$^2$~$^2$D$_{3/2}$,   
8; triangles: 3s3p$^2$~$^2$P$_{1/2}$,  9; stars: 3s3p$^2$~$^2$S$_{1/2}$,  10; squares: 3s3p$^2$~$^2$P$_{3/2}$,  and 11; diamonds: 3s$^2$3d~$^2$D$_{3/2}$) of As~XXI. Corresponding data in (b) are from the level 2 (3s$^2$3p~$^2$P$^o_{3/2}$). In (c) and (d) are our collision strengths with FAC and in (e) and (f) the effective collision strengths for the same corresponding transitions.}
 \end{figure*}
 
 \clearpage

\setcounter{figure}{3}
\begin{figure*}
\includegraphics[angle=-90,width=0.8\textwidth]{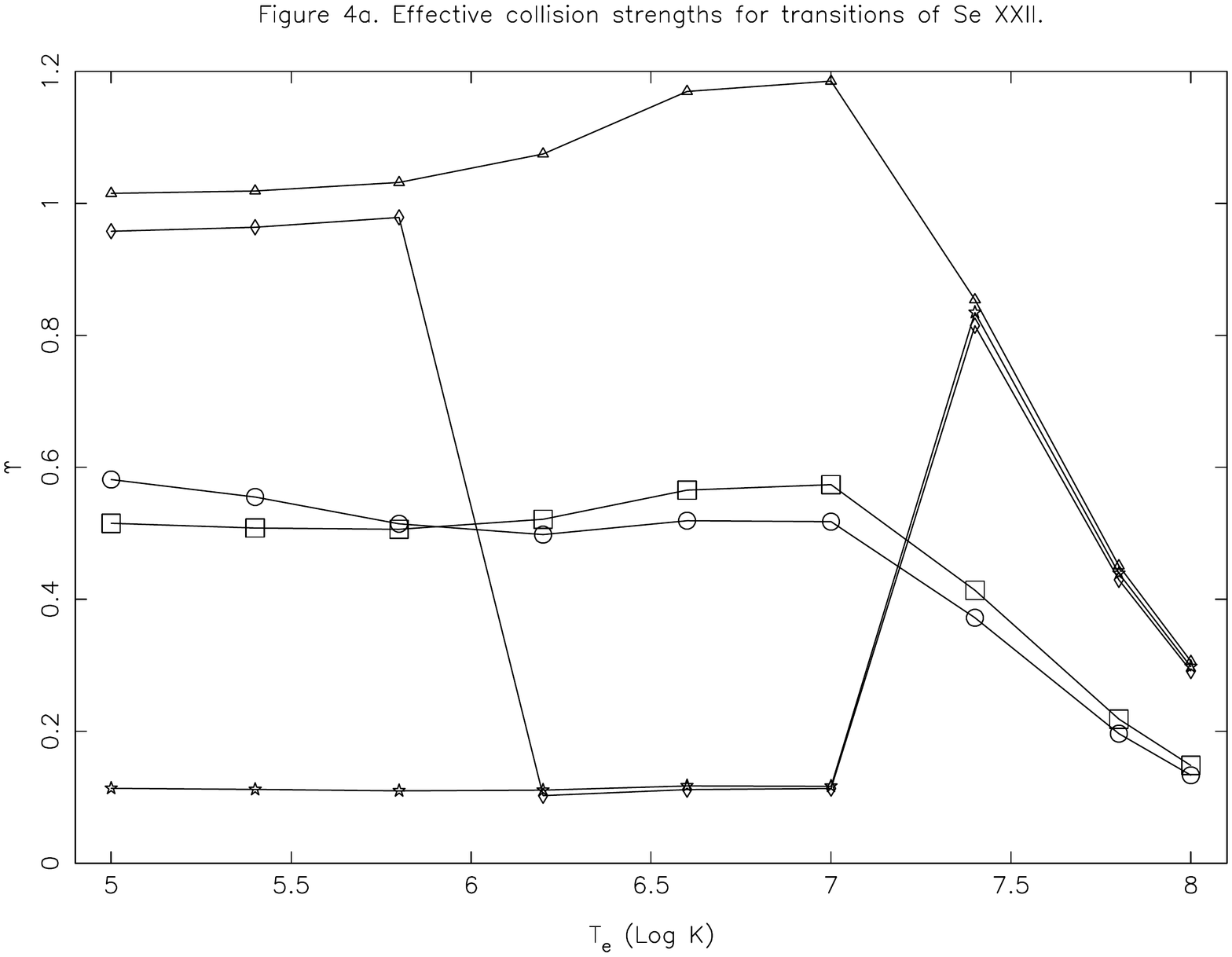}
 \vspace{-1.5cm}
 \caption{.}
 \end{figure*}

\setcounter{figure}{3}
 \begin{figure*}
\includegraphics[angle=-90,width=0.8\textwidth]{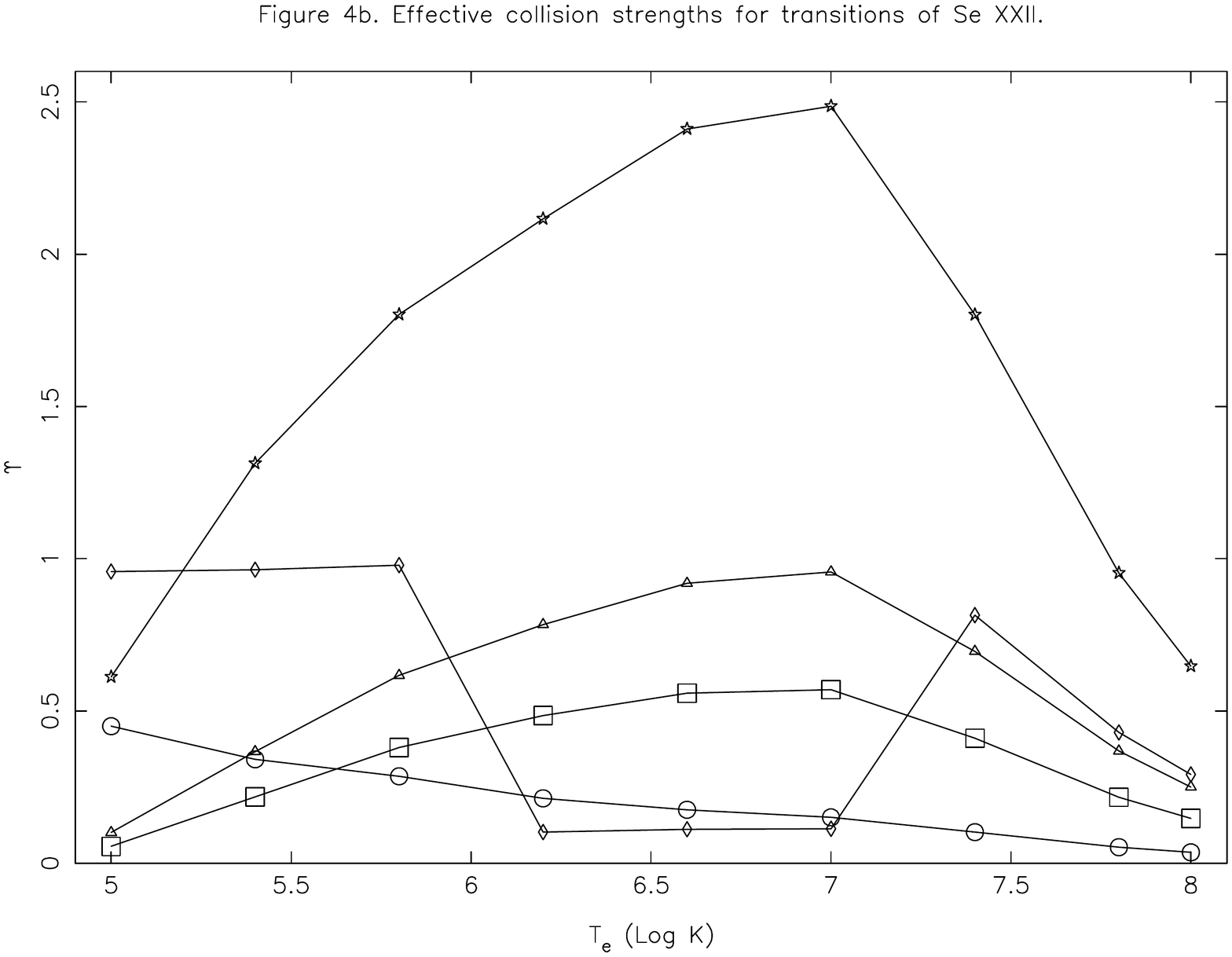}
 \vspace{-1.5cm}
\caption{}
 \end{figure*}

\setcounter{figure}{3}

 \begin{figure*}
\includegraphics[angle=-90,width=0.8\textwidth]{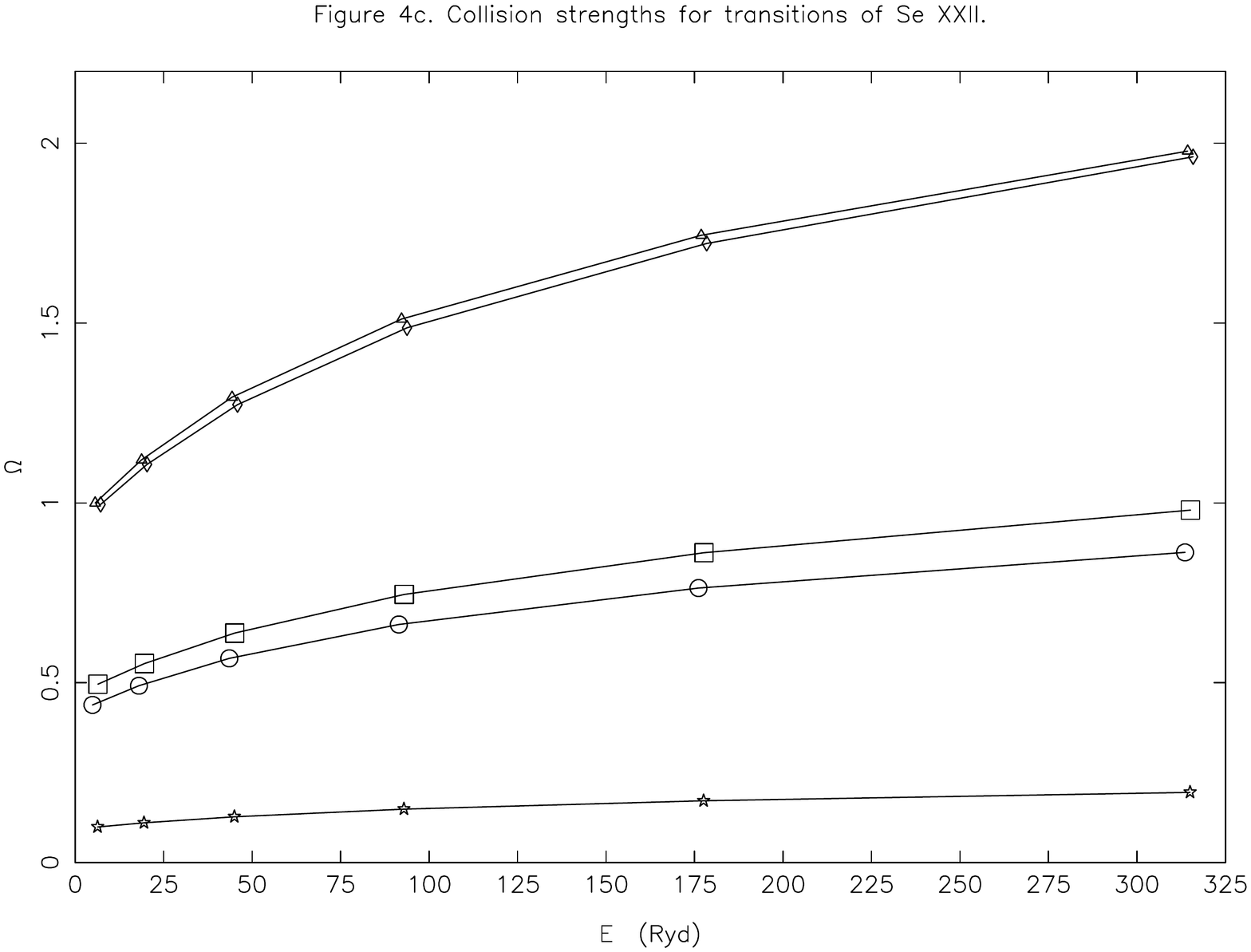}
 \vspace{-1.5cm}
\caption{}
 \end{figure*}

\setcounter{figure}{3}
 \begin{figure*}
\includegraphics[angle=-90,width=0.8\textwidth]{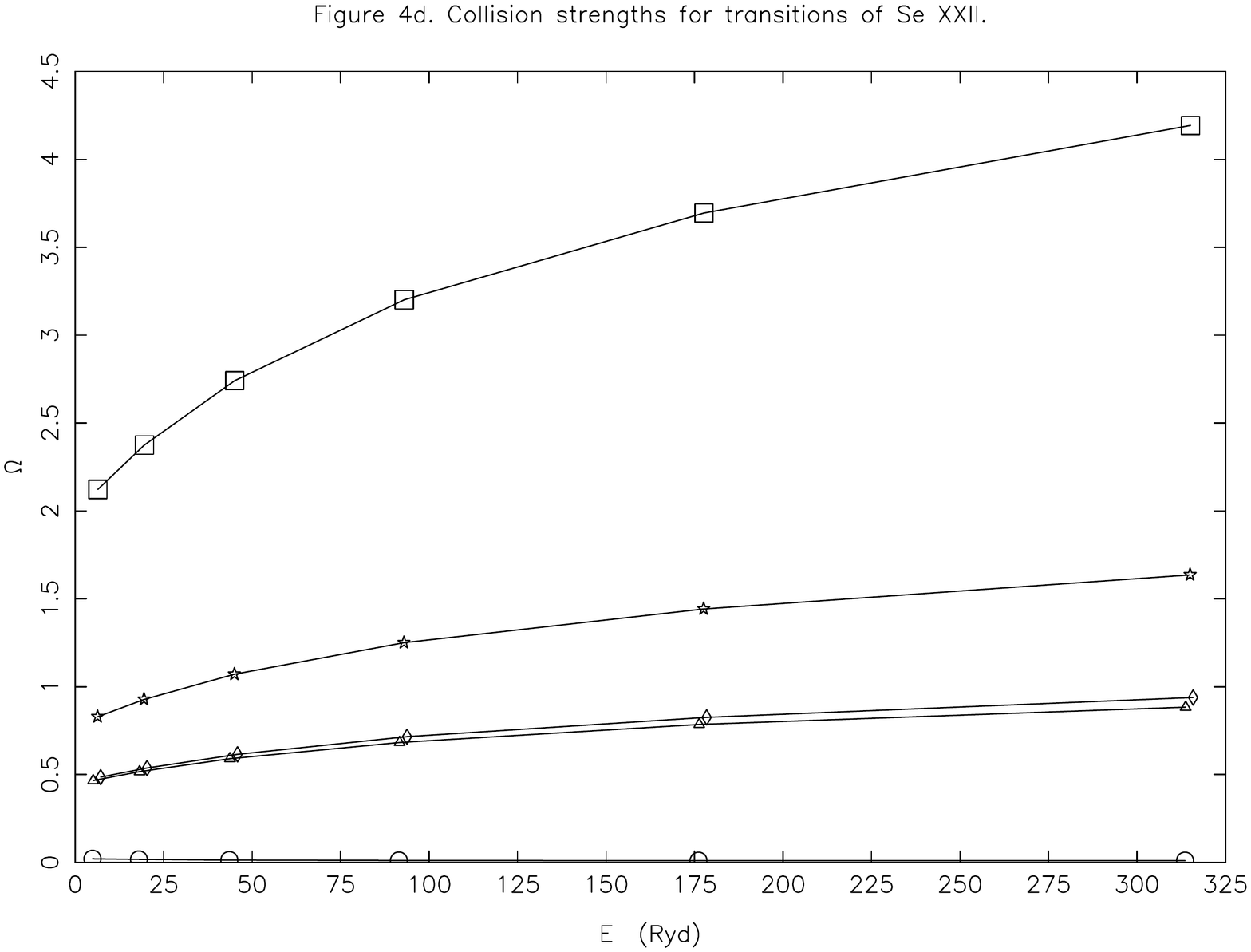}
 \vspace{-1.5cm}
\caption{}
 \end{figure*}

\setcounter{figure}{3}
 \begin{figure*}
\includegraphics[angle=-90,width=0.8\textwidth]{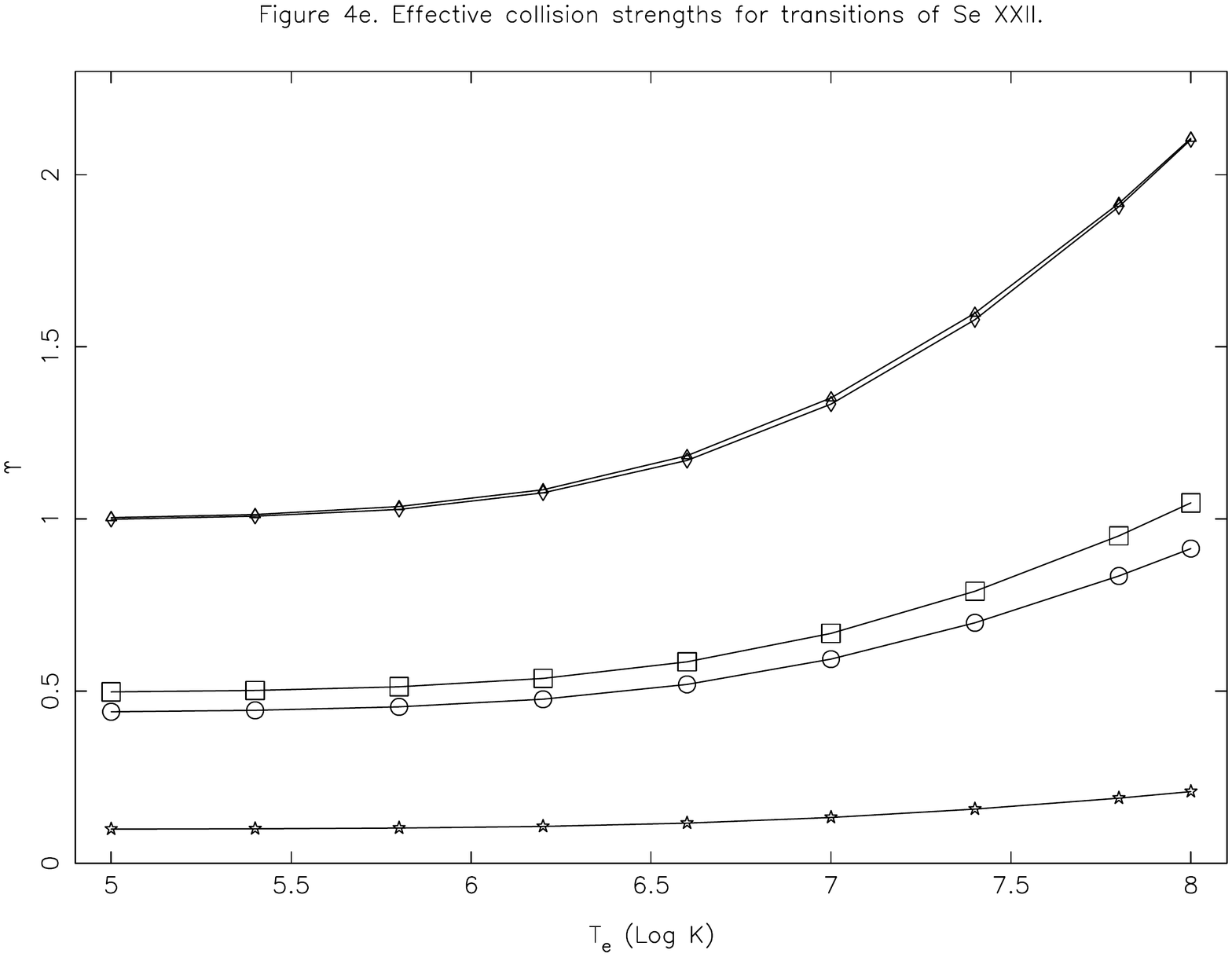}
 \vspace{-1.5cm}
\caption{}
 \end{figure*}

\setcounter{figure}{3}
 \begin{figure*}
\includegraphics[angle=-90,width=0.8\textwidth]{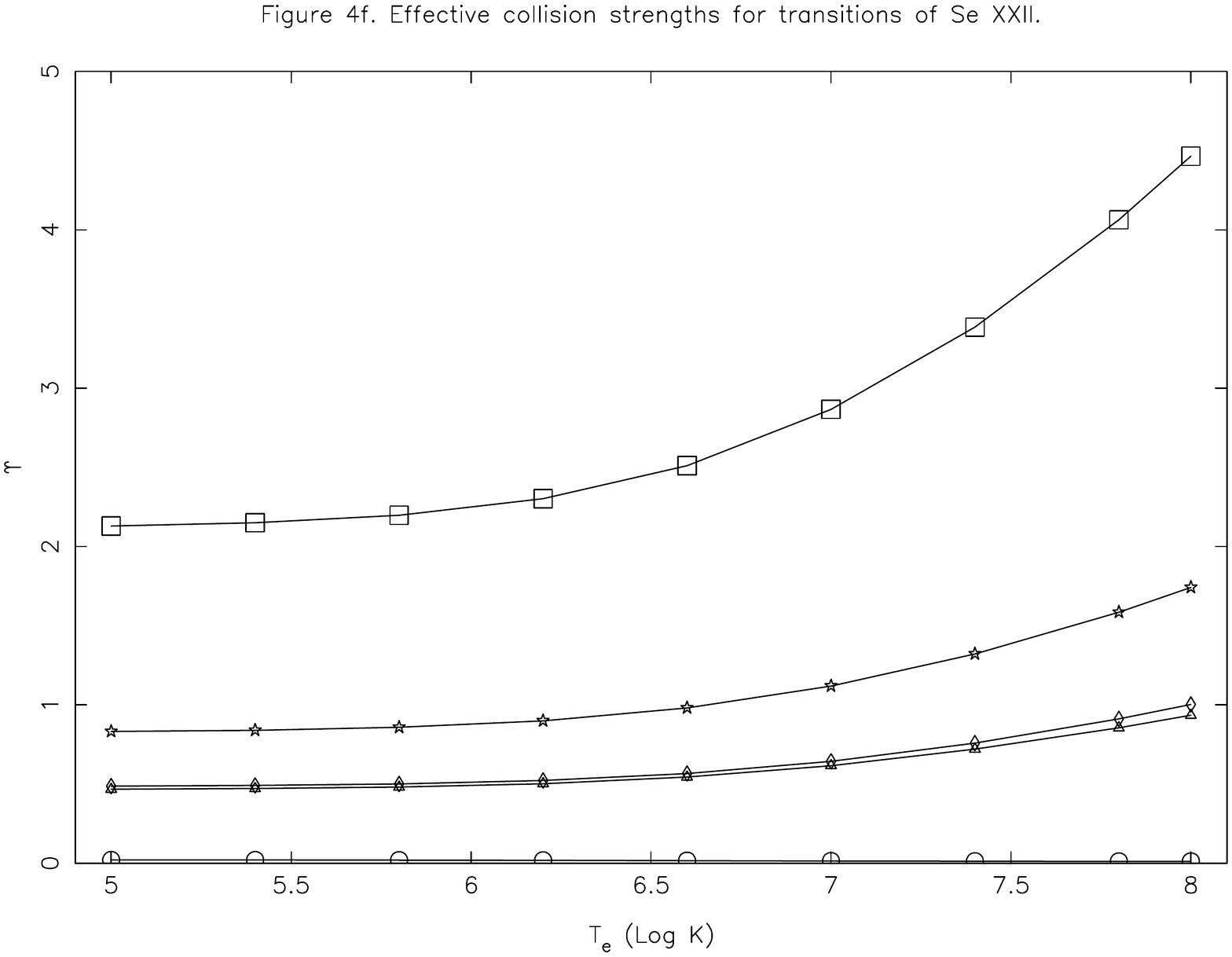}
 \vspace{-1.5cm}
\caption{Effective collision strengths of \cite{wj} (a):  from the ground level (1: 3s$^2$3p~$^2$P$^o_{1/2}$) to the excited levels (6; circles: 3s3p$^2$~$^2$D$_{3/2}$,   
8; triangles: 3s3p$^2$~$^2$P$_{1/2}$,  9; stars: 3s3p$^2$~$^2$S$_{1/2}$,  10; squares: 3s3p$^2$~$^2$P$_{3/2}$,  and 11; diamonds: 3s$^2$3d~$^2$D$_{3/2}$) of Se~XXII. Corresponding data in (b) are from the level 2 (3s$^2$3p~$^2$P$^o_{3/2}$). In (c) and (d) are our collision strengths with FAC and in (e) and (f) the effective collision strengths for the same corresponding transitions.}
 \end{figure*}
 
\setcounter{figure}{4}
\begin{figure*}
\includegraphics[angle=-90,width=0.8\textwidth]{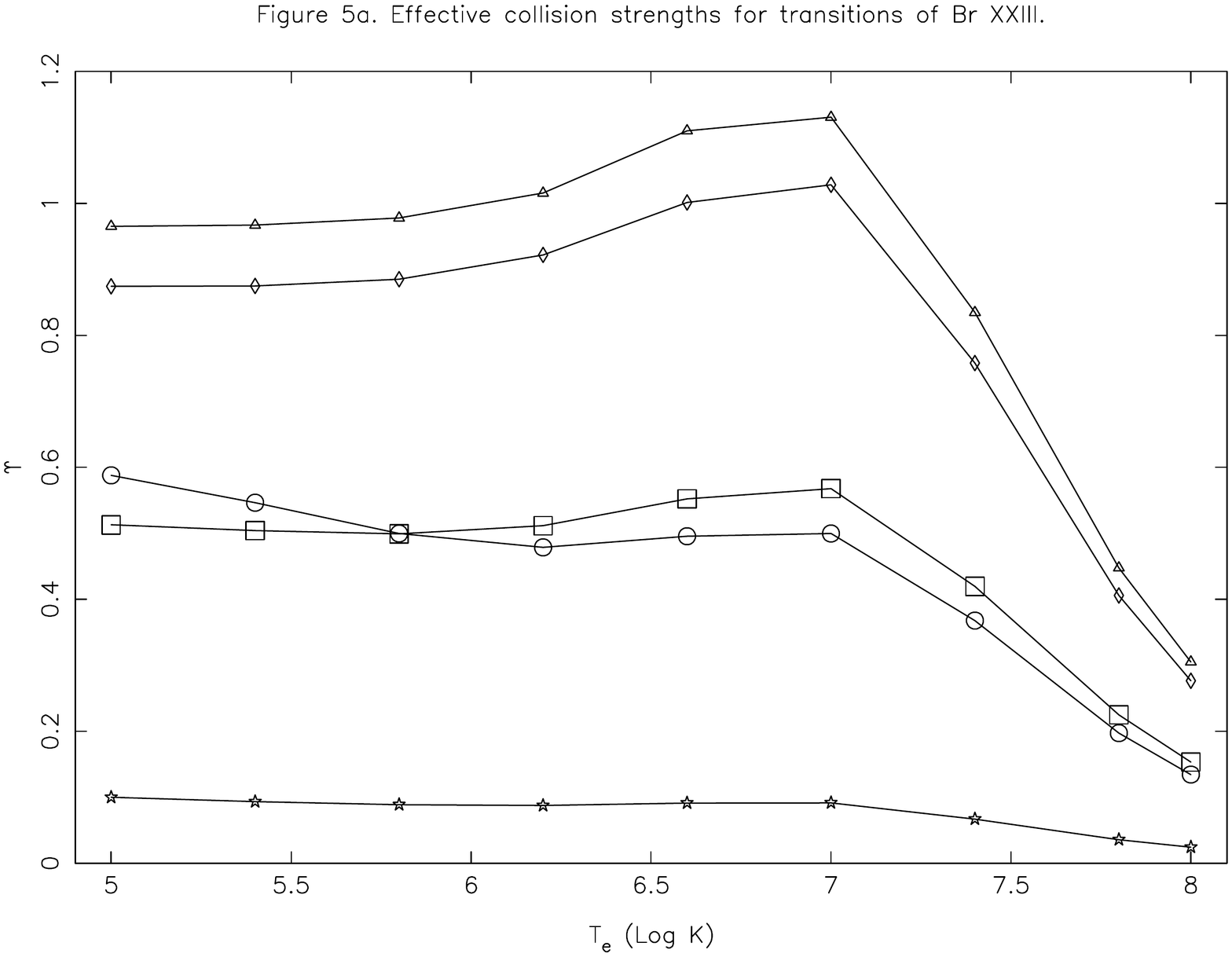}
 \vspace{-1.5cm}
 \caption{}
 \end{figure*}

\setcounter{figure}{4}
 \begin{figure*}
\includegraphics[angle=-90,width=0.8\textwidth]{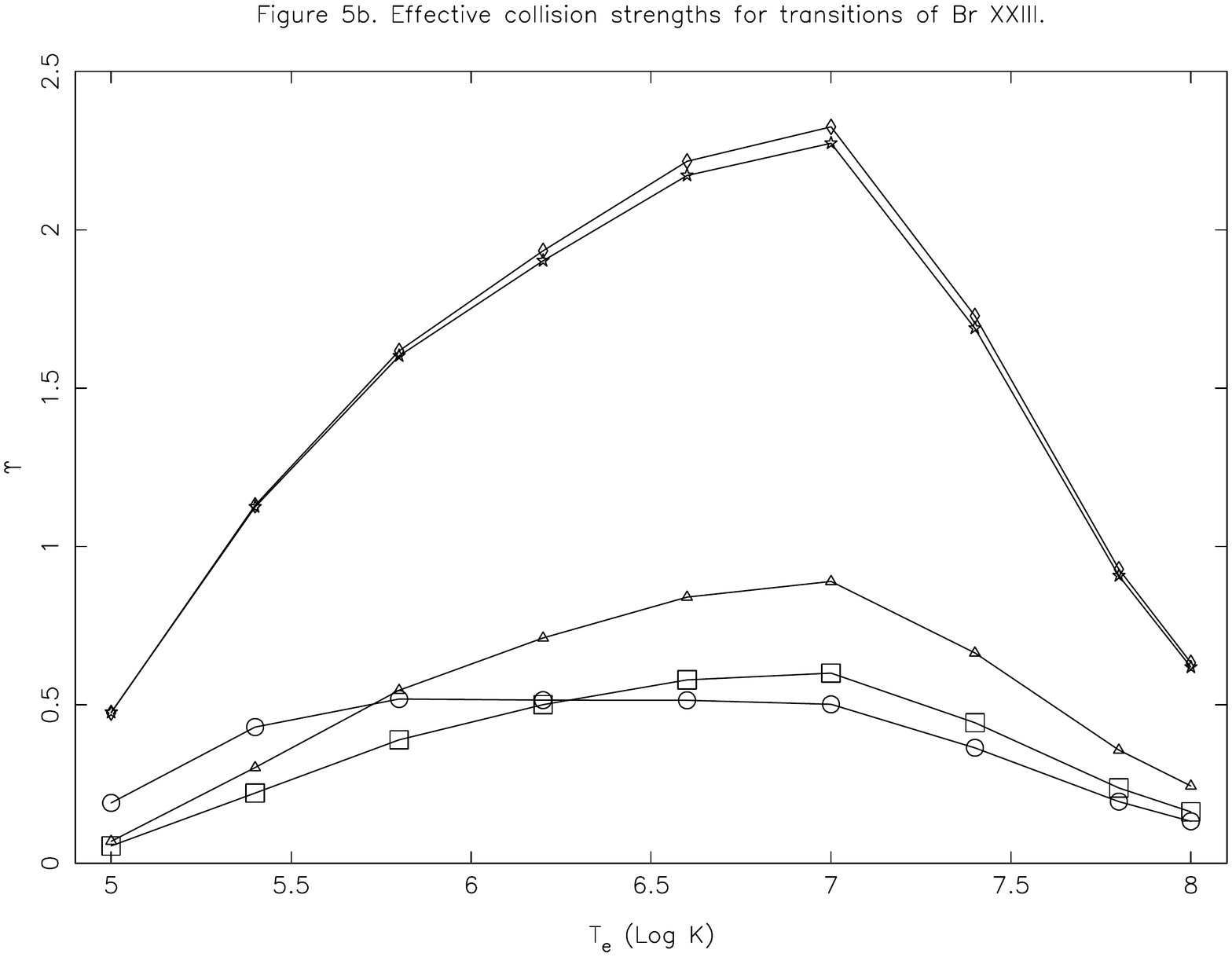}
 \vspace{-1.5cm}
\caption{}
 \end{figure*}

\setcounter{figure}{4}

 \begin{figure*}
\includegraphics[angle=-90,width=0.8\textwidth]{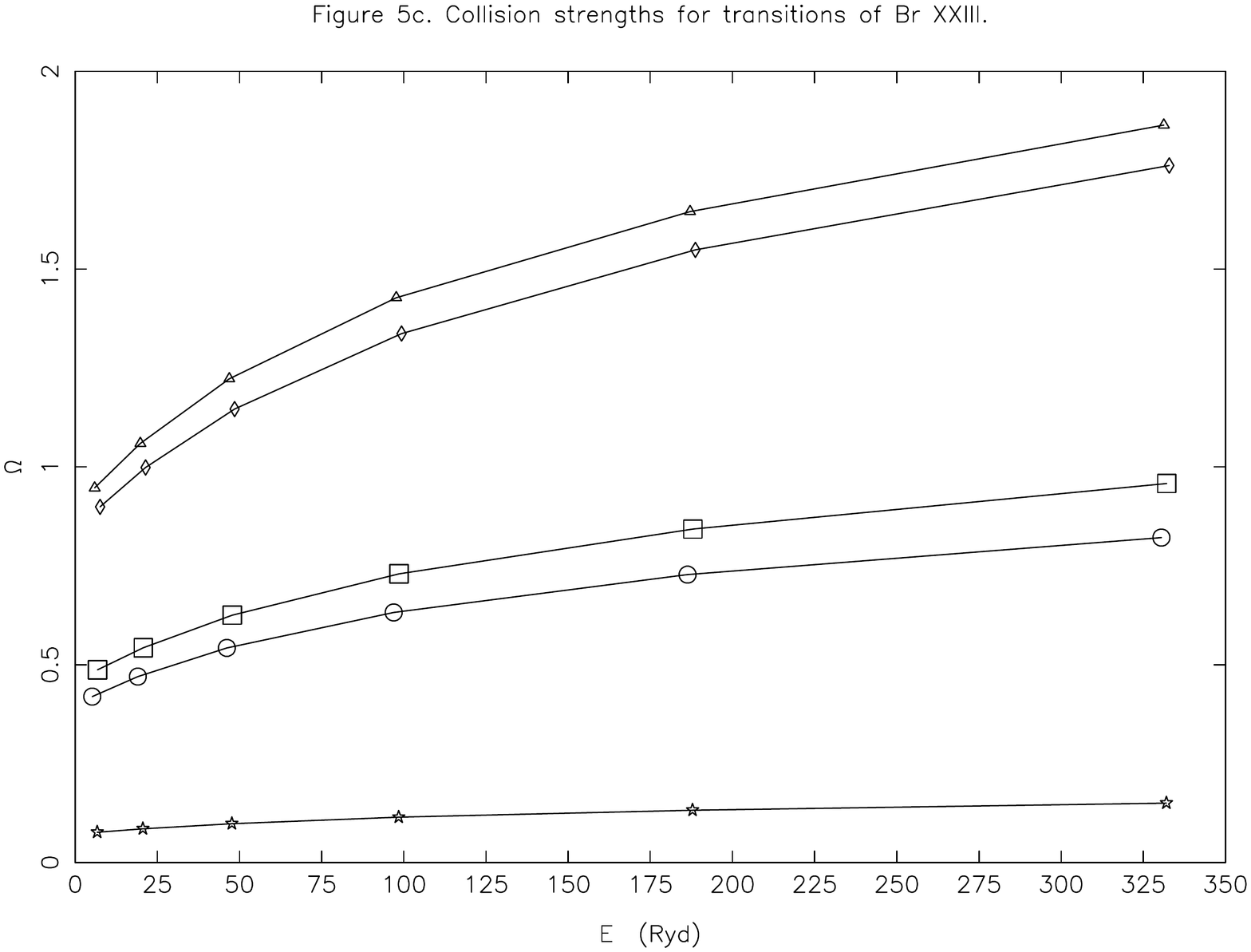}
 \vspace{-1.5cm}
\caption{}
 \end{figure*}

\setcounter{figure}{4}
 \begin{figure*}
\includegraphics[angle=-90,width=0.8\textwidth]{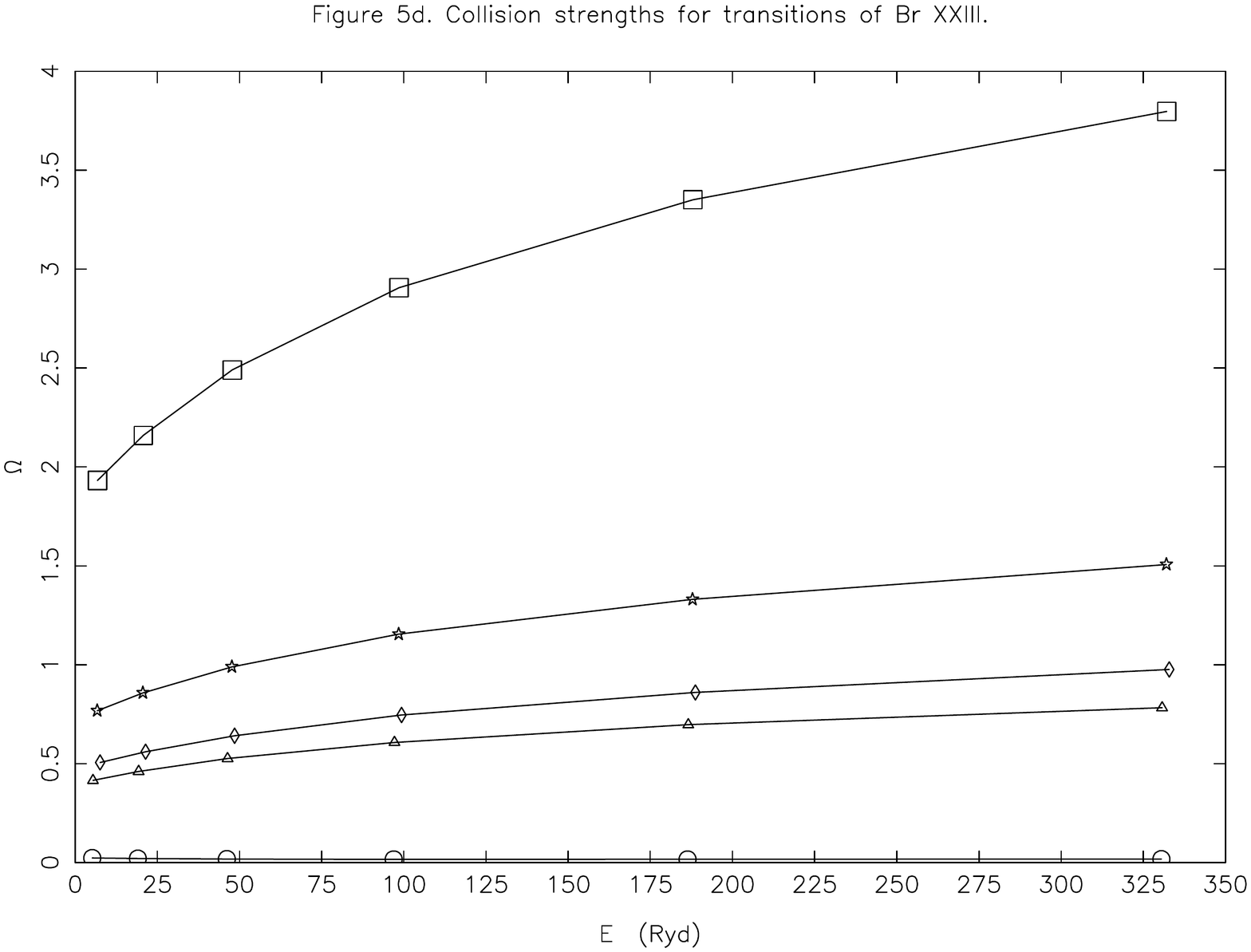}
 \vspace{-1.5cm}
\caption{}
 \end{figure*}

\setcounter{figure}{4}
 \begin{figure*}
\includegraphics[angle=-90,width=0.8\textwidth]{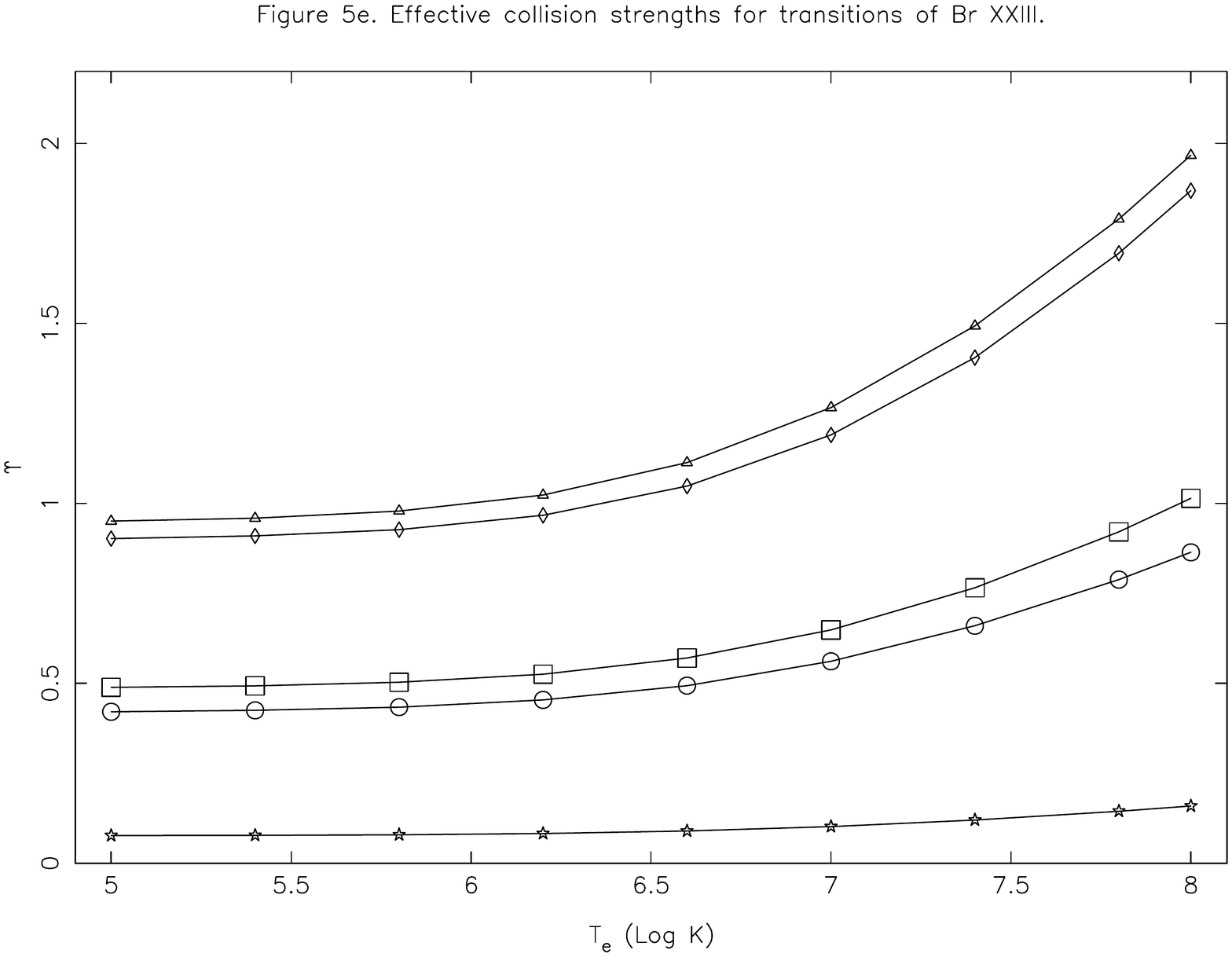}
 \vspace{-1.5cm}
\caption{}
 \end{figure*}

\setcounter{figure}{4}
 \begin{figure*}
\includegraphics[angle=-90,width=0.8\textwidth]{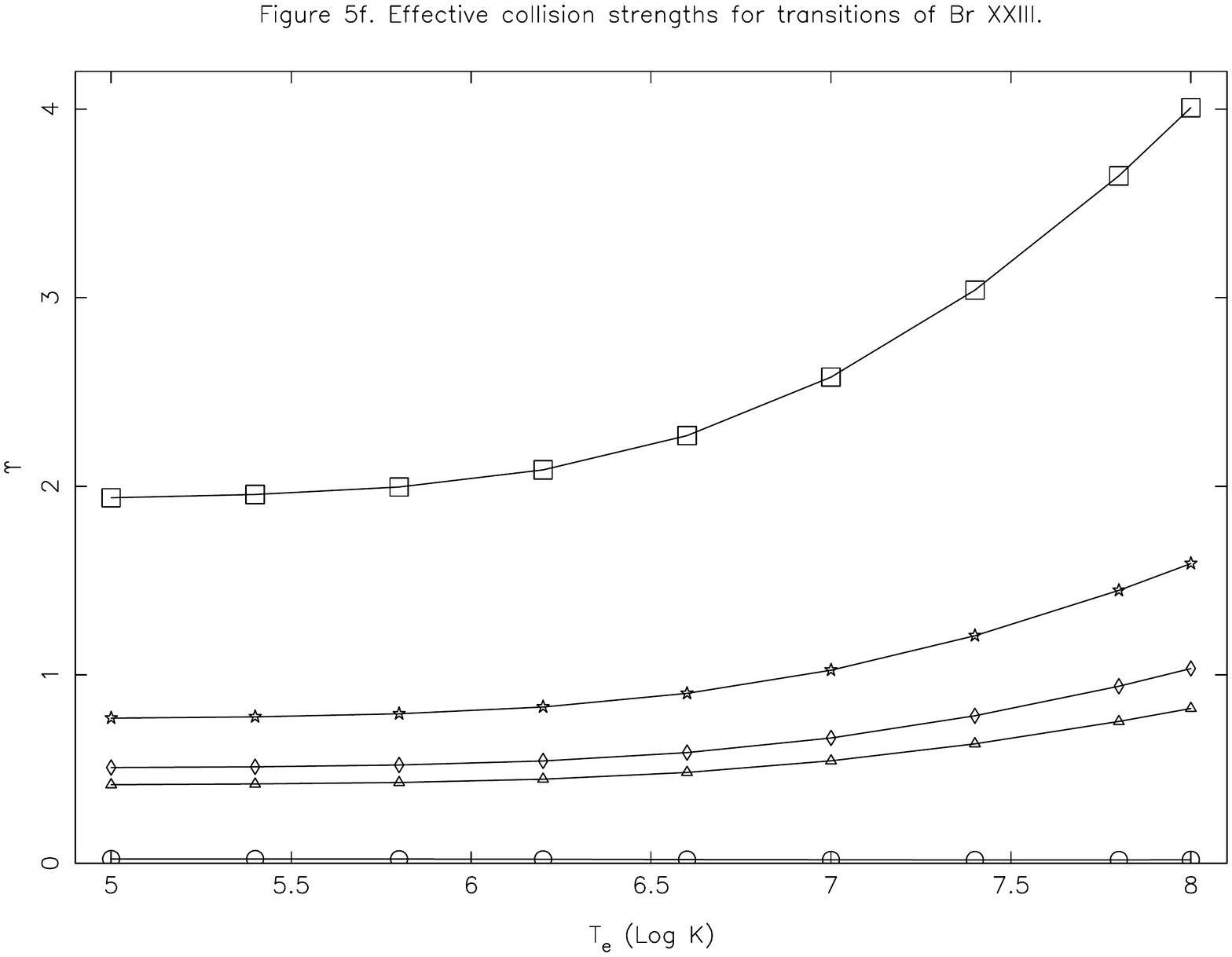}
 \vspace{-1.5cm}
\caption{Effective collision strengths of \cite{wj} (a):  from the ground level (1: 3s$^2$3p~$^2$P$^o_{1/2}$) to the excited levels (6; circles: 3s3p$^2$~$^2$D$_{3/2}$,   
8; triangles: 3s3p$^2$~$^2$P$_{1/2}$,  9; stars: 3s3p$^2$~$^2$S$_{1/2}$,  10; squares: 3s3p$^2$~$^2$P$_{3/2}$,  and 11; diamonds: 3s$^2$3d~$^2$D$_{3/2}$) of Br~XXIII. Corresponding data in (b) are from the level 2 (3s$^2$3p~$^2$P$^o_{3/2}$). In (c) and (d) are our collision strengths with FAC and in (e) and (f) the effective collision strengths for the same corresponding transitions.}
 \end{figure*}

 \newpage
 


\clearpage

\renewcommand{\baselinestretch}{1.5}

\footnotesize  

\begin{longtable}{rlrrrrrrrrrrrrrrr}
\caption{\label{table_A}
Threshold energies (in Ryd) of the lowest 30 levels of  Ga XIX, Ge XX, As XXI, Se XXII, and Br XXIII.}
}
{Index & 
\multicolumn{1}{l}{Configuration} & 
\multicolumn{1 }{l}{Level} & 
\multicolumn{2}{c}{Ga~XIX} &
\multicolumn{2}{c}{Ge~XX} &
\multicolumn{2}{c}{As~XXI} &
\multicolumn{2}{c}{Se~XXII} &
\multicolumn{2}{c}{Br~XXIII} \\
\hline
& & & \multicolumn{1}{r}{GRASP} &
\multicolumn{1}{r}{FAC} &
\multicolumn{1}{r}{GRASP} &
\multicolumn{1}{r}{FAC} &
\multicolumn{1}{r}{GRASP} &
\multicolumn{1}{r}{FAC} &
\multicolumn{1}{r}{GRASP} &
\multicolumn{1}{r}{FAC} &
\multicolumn{1}{r}{GRASP} &
\multicolumn{1}{r}{FAC} \\
\hline
\endfirsthead
\caption[]{(continued)}
Index & 
\multicolumn{1}{l}{Configuration} & 
\multicolumn{1 }{l}{Level} & 
\multicolumn{2}{c}{Ga~XIX} &
\multicolumn{2}{c}{Ge~XX} &
\multicolumn{2}{c}{As~XXI} &
\multicolumn{2}{c}{Se~XXII} &
\multicolumn{2}{c}{Br~XXIII} \\
\hline
& & & \multicolumn{1}{r}{GRASP} &
\multicolumn{1}{r}{FAC} &
\multicolumn{1}{r}{GRASP} &
\multicolumn{1}{r}{FAC} &
\multicolumn{1}{r}{GRASP} &
\multicolumn{1}{r}{FAC} &
\multicolumn{1}{r}{GRASP} &
\multicolumn{1}{r}{FAC} &
\multicolumn{1}{r}{GRASP} &
\multicolumn{1}{r}{FAC} \\

\hline
\endhead
    1 & 3s$^2$3p	    &	  $^2$P$^o_{1/2 }$  & 0.0000	&   0.0000   &  0.0000  & 0.0000   &   0.0000  &  0.0000   &   0.0000  &   0.0000  &   0.0000  &  0.0000   \\
    2 & 3s$^2$3p	    &	  $^2$P$^o_{3/2 }$  & 0.4221	&   0.4217   &  0.4948  & 0.4943   &   0.5765  &  0.5761   &   0.6680  &   0.6676  &   0.7701  &  0.7696   \\
    3 & 3s3p$^2$	    &	  $^4$P$  _{1/2 }$  & 2.8212	&   2.8202   &  2.9938  & 2.9923   &   3.1699  &  3.1680   &   3.3495  &   3.3471  &   3.5325  &  3.5295   \\
    4 & 3s3p$^2$	    &	  $^4$P$  _{3/2 }$  & 3.0152	&   3.0142   &  3.2275  & 3.2262   &   3.4500  &  3.4484   &   3.6836  &   3.6815  &   3.9288  &  3.9264   \\
    5 & 3s3p$^2$	    &	  $^4$P$  _{5/2 }$  & 3.2090	&   3.2076   &  3.4479  & 3.4461   &   3.6982  &  3.6960   &   3.9606  &   3.9580  &   4.2356  &  4.2325   \\
    6 & 3s3p$^2$	    &	  $^2$D$  _{3/2 }$  & 3.8688	&   3.8660   &  4.1269  & 4.1237   &   4.3959  &  4.3924   &   4.6766  &   4.6727  &   4.9697  &  4.9654   \\
    7 & 3s3p$^2$	    &	  $^2$D$  _{5/2 }$  & 3.9529	&   3.9501   &  4.2360  & 4.2329   &   4.5361  &  4.5327   &   4.8548  &   4.8512  &   5.1940  &  5.1901   \\
    8 & 3s3p$^2$	    &	  $^2$P$  _{1/2 }$  & 4.6115	&   4.6012   &  4.8803  & 4.8695   &   5.1571  &  5.1458   &   5.4427  &   5.4308  &   5.7380  &  5.7257   \\
    9 & 3s3p$^2$	    &	  $^2$S$  _{1/2 }$  & 5.0439	&   5.0331   &  5.3797  & 5.3686   &   5.7338  &  5.7225   &   6.1078  &   6.0962  &   6.5034  &  6.4914   \\
   10 & 3s3p$^2$	    &	  $^2$P$  _{3/2 }$  & 5.1565	&   5.1447   &  5.4956  & 5.4835   &   5.8511  &  5.8386   &   6.2242  &   6.2114  &   6.6160  &  6.6030   \\
   11 & 3s$^2$3d	    &	  $^2$D$  _{3/2 }$  & 5.9338	&   5.9226   &  6.2714  & 6.2601   &   6.6192  &  6.6080   &   6.9786  &   6.9675  &   7.3511  &  7.3399   \\
   12 & 3s$^2$3d	    &	  $^2$D$  _{5/2 }$  & 5.9864	&   5.9750   &  6.3345  & 6.3232   &   6.6939  &  6.6827   &   7.0659  &   7.0548  &   7.4515  &  7.4405   \\
   13 & 3p$^3$  	    &	  $^2$P$^o_{3/2 }$  & 7.3254	&   7.3192   &  7.7661  & 7.7590   &   8.2175  &  8.2094   &   8.6801  &   8.6711  &   9.1547  &  9.1448   \\
   14 & 3p$^3$  	    &	  $^2$D$^o_{5/2 }$  & 7.4473	&   7.4421   &  7.9232  & 7.9175   &   8.4171  &  8.4108   &   8.9302  &   8.9233  &   9.4636  &  9.4561   \\
   15 & 3p$^3$  	    &	  $^2$D$^o_{3/2 }$  & 7.6065	&   7.5949   &  8.0826  & 8.0705   &   8.5774  &  8.5649   &   9.0922  &   9.0792  &   9.6277  &  9.6143   \\
   16 & 3s3p3d	            &	  $^2$P$^o_{3/2 }$  & 8.0069	&   8.0028   &  8.4650  & 8.4605   &   8.9338  &  8.9290   &   9.4146  &   9.4094  &   9.9089  &  9.9033   \\
   17 & 3s3p3d  	    &	  $^2$F$^o_{5/2 }$  & 8.0959	&   8.0919   &  8.5686  & 8.5643   &   9.0535  &  9.0489   &   9.5516  &   9.5467  &  10.0641  & 10.0589   \\
   18 & 3s3p3d  	    &	  $^2$F$^o_{7/2 }$  & 8.2303	&   8.2265   &  8.7260  & 8.7219   &   9.2363  &  9.2320   &   9.7624  &   9.7577  &  10.3051  & 10.3001   \\
   19 & 3p$^3$  	    &	  $^2$P$^o_{1/2 }$  & 8.3275	&   8.3150   &  8.8452  & 8.8322   &   9.3826  &  9.3690   &   9.9412  &   9.9270  &  10.5227  & 10.5078   \\
   20 & 3p$^3$  	    &	  $^4$S$^o_{3/2 }$  & 8.4356	&   8.4241   &  8.9694  & 8.9654   &   9.5260  &  9.5219   &  10.1054  &  10.1010  &  10.6879  & 10.6790   \\
   21 & 3s3p3d  	    &	  $^4$F$^o_{9/2 }$  & 8.4337	&   8.4299   &  8.9831  & 8.9712   &   9.5523  &  9.5403   &  10.1337  &  10.1221  &  10.7004  & 10.6900   \\
   22 & 3s3p3d  	    &	  $^2$D$^o_{5/2 }$  & 8.6409	&   8.6330   &  9.1331  & 9.1250   &   9.6376  &  9.6292   &  10.1554  &  10.1468  &  10.7093  & 10.7047   \\
   23 & 3s3p3d  	    &	  $^2$D$^o_{3/2 }$  & 8.6803	&   8.6718   &  9.1774  & 9.1686   &   9.6905  &  9.6813   &  10.2248  &  10.2151  &  10.7592  & 10.7491   \\
   24 & 3s3p3d  	    &	  $^2$P$^o_{1/2 }$  & 8.7013	&   8.6921   &  9.1968  & 9.1875   &   9.7044  &  9.6948   &  10.2318  &  10.2216  &  10.8313  & 10.8191   \\
   25 & 3s3p3d  	    &	  $^4$F$^o_{7/2 }$  & 8.9310	&   8.9217   &  9.4721  & 9.4627   &  10.0416  & 10.0230   &  10.6139  &  10.6043  &  11.2181  & 11.2083   \\
   26 & 3s3p3d  	    &	  $^4$P$^o_{1/2 }$  & 8.9340	&   8.9270   &  9.4784  & 9.4712   &  10.0326  & 10.0343   &  10.6253  &  10.6178  &  11.2312  & 11.2236   \\
   27 & 3s3p3d  	    &	  $^4$D$^o_{5/2 }$  & 8.9436	&   8.9349   &  9.4864  & 9.4774   &  10.0479  & 10.0388   &  10.6298  &  10.6204  &  11.2337  & 11.2241   \\
   28 & 3s3p3d  	    &	  $^4$D$^o_{3/2 }$  & 8.9430	&   8.9352   &  9.4876  & 9.4796   &  10.0512  & 10.0431   &  10.6357  &  10.6274  &  11.2427  & 11.2342   \\
   29 & 3s3p3d  	    &	  $^4$P$^o_{5/2 }$  & 9.2139	&   9.1985   &  9.7547  & 9.7387   &  10.3074  & 10.2910   &  10.8715  &  10.8544  &  11.4469  & 11.4292   \\
   30 & 3s3p3d  	    &	  $^2$D$^o_{3/2 }$  & 9.2157	&   9.2003   &  9.7625  & 9.7469   &  10.3273  & 10.3113   &  10.9112  &  10.8948  &  11.5156  & 11.4987   \\
\hline               
\end{longtable}

\begin{flushleft}
GRASP: Earlier calculations of \cite{wj} with the {\sc grasp} code  \\ 
FAC: Present calculations with the {\sc fac} code   \\   
\end{flushleft}


\newpage
\begin{longtable}{rrlllrrlll}
\caption{\label{table_B}
Transition rates (A, s$^{-1}$) and oscillator strengths (f) from the lowest two levels of Al-like ions. $a{\pm}b \equiv a{\times}$10$^{{\pm}b}$. \\See Table A for level indices.\\}
} 
{I &
\multicolumn{1}{r}{J} &
\multicolumn{1}{r}{A(G)} &
\multicolumn{1}{r}{A(F)} &
\multicolumn{1}{r}{f(F)} &
I &
\multicolumn{1}{r}{J} &
\multicolumn{1}{r}{A(G)} &
\multicolumn{1}{r}{A(F)} &
\multicolumn{1}{r}{f(F)} \\
\hline
\endfirsthead\\
\caption[]{(continued)}
 I &
\multicolumn{1}{r}{J} &
\multicolumn{1}{r}{Type} &
\multicolumn{1}{r}{GRASP1} &
\multicolumn{1}{r}{GRASP2} &
\multicolumn{1}{r}{MCHF} \\
\hline
\endhead
{\bf Ga~XIX} \\ 
    1  &  6 & 4.80+09 & 4.85+09 & 8.07-2 & 2  &  6  & 3.63+06 &  3.52+06 & 3.70-5  \\ 
    1  &  8 & 4.02+10 & 3.99+10 & 2.34-1 & 2  &  7  & 2.88+09 &  2.90+09 & 4.35-2  \\ 
    1  &  9 & 1.05+10 & 1.02+10 & 5.02-2 & 2  &  9  & 3.67+10 &  3.64+10 & 1.06-1  \\ 
    1  & 10 & 1.45+10 & 1.42+10 & 1.34-1 & 2  & 10  & 5.24+10 &  5.18+10 & 2.89-1  \\ 
    1  & 11 & 5.39+10 & 5.31+10 & 3.77-1 & 2  & 11  & 1.45+10 &  1.42+10 & 5.83-2  \\ 
{\bf Ge~XX} \\   
    1  &  6 & 5.52+09 &  5.58+09 &  8.17-2 & 2  &  6  & 9.48+04 &	    &	       \\
    1  &  8 & 4.52+10 &  4.49+10 &  2.36-1 & 2  &  7  & 3.07+09 &   3.10+09 &  4.14-2  \\
    1  &  9 & 9.69+09 &  9.39+09 &  4.06-2 & 2  &  9  & 3.98+10 &   3.94+10 &  1.03-1  \\
    1  & 10 & 1.65=10 &  1.62+10 &  1.34-1 & 2  & 10  & 5.61=10 &   5.56+10 &  2.78-1  \\
    1  & 11 & 5.71+10 &  5.64+10 &  3.58-1 & 2  & 11  & 1.64+10 &   1.60+10 &  6.00-2  \\
{\bf As XXI} \\    
    1  &  6 & 6.34+09 &  6.41+09 &  8.27-2 & 2  &  6  & 7.80+06 &   8.04+06 &  6.87-5  \\
    1  &  8 & 5.05+10 &  5.00+10 &  2.35-1 & 2  &  7  & 3.25+09 &   3.28+09 &  3.92-2  \\
    1  &  9 & 8.94+09 &  8.66+09 &  3.29-2 & 2  &  9  & 4.30+10 &   4.25+10 &  9.99-2  \\
    1  & 10 & 1.90+10 &  1.85+10 &  1.35-1 & 2  & 10  & 5.98+10 &   5.93+10 &  2.67-1  \\
    1  & 11 & 6.04+10 &  5.97+10 &  3.40-1 & 2  & 11  & 1.88+10 &   1.83+10 &  6.27-2  \\    
{\bf Se XXII }  \\ 
    1  &  6 & 7.27+09 &  7.35+09 &  8.38-2 & 2  &  6  & 3.07+07 &   3.12+07 &  2.42-4  \\    
    1  &  8 & 5.59+10 &  5.54+10 &  2.34-1 & 2  &  7  & 3.42+09 &   3.46+09 &  3.69-2  \\
    1  &  9 & 8.28+09 &  8.02+09 &  2.69-2 & 2  &  9  & 4.63+10 &   4.58+10 &  9.68-2  \\    
    1  & 10 & 2.21+10 &  2.15+10 &  1.39-1 & 2  & 10  & 6.33+10 &   6.29+10 &  2.55-1  \\    
    1  & 11 & 6.34+10 &  6.28+10 &  3.22-1 & 2  & 11  & 2.18+10 &   2.12+10 &  6.65-2  \\  
{\bf Br XXIII} \\    
    1  &  6 & 8.31+09 &  8.41+09 &  8.50-2 & 2  &  6  & 7.29+07 &   7.38+07 &  5.22-4  \\   
    1  &  8 & 6.17+10 &  6.11+10 &  2.32-1 & 2  &  7  & 3.57+09 &   3.61+09 &  3.45-2  \\    
    1  &  9 & 7.70+09 &  7.46+09 &  2.20-2 & 2  &  9  & 4.98+10 &   4.93+10 &  9.37-2  \\    
    1  & 10 & 2.60+10 &  2.52+10 &  1.44-1 & 2  & 10  & 6.65+10 &   6.62+10 &  2.42-1  \\    
    1  & 11 & 6.51+10 &  6.56+10 &  3.03-1 & 2  & 11  & 2.57+10 &   2.49+10 &  7.17-2  \\

\hline   
\end{longtable}
\begin{flushleft}
A(G): Earlier A-values  of \cite{wj} with the {\sc grasp} code  \\ 
A(F): Present A-values with the {\sc fac} code  \\ 
f(F): Present f-values  with the {\sc fac} code 
\end{flushleft}


\section*{References}
\begin{thebibliography}{999}
\bibitem[\protect\citeauthoryear{Wang \& Jiang}{2022}]{wj} 
H.B. Wang, G. Jiang. At. Data Nucl. Data Tables 148 (2022) 101532.
\bibitem[\protect\citeauthoryear{Grant et al.}{1980}]{grasp0}
I.P. Grant, B.J. McKenzie,  P.H. Norrington,  D.F. Mayers,   N.C. Pyper N. C.    Comput. Phys. Commun.  21 (1980)  207.  
\bibitem[\protect\citeauthoryear{Jonsson et al.}{2023}]{grasp}
P. J{\"o}nsson, G. Gaigalas, C. Froese  Fischer, J. Bier{\'o}n, I.P. Grant, T. Brage, J. Ekman, M. Godefroid, J. Grumer, J. Li, W. Li. Atoms 11 (2023) 68.
\bibitem{fac}
M.F. Gu,  Can. J.  Phys, 86 (2008) 675.
\bibitem{nlike}
H.B. Wang, G. Jiang, X.F. Li, Z.C. He. At. Data Nucl. Data Tables 120 (2018) 373.
\bibitem{km1}
K.M. Aggarwal. At. Data Nucl. Data Tables 120 (2018) 430.
\bibitem{km2} 
K.M. Aggarwal, F.P. Keenan. Fusion. Sci. Tech. 63 (2013) 363.
\bibitem{km3}
K.M. Aggarwal. Atoms 5 (2017) 37.





\end{thebibliography}
\end{document}